\documentclass[a4paper,12pt]{article}

\usepackage{amsmath,amssymb,latexsym}
\usepackage[dvips]{graphicx}

\newcommand{\ev}{\mathrm{ev}}
\newcommand{\coev}{\mathrm{coev}}
\newcommand{\diff}{\mathrm{diff}}
\newcommand{\dd}[1]{\frac{\delta}{\delta \phi(#1)}}
\newcommand{\id}{\mathrm{id}}
\newcommand{\X}{X^{\ast}}
\newcommand{\A}{\mathcal{A}}
\newcommand{\1}{\mathbf{1}}
\newcommand{\Xh}{\widehat{X}}
\newcommand{\otimesh}{\hat{\otimes}}

\newcommand{\Zfreed}[1]{Z^{(0)}_{#1}}
\newcommand{\phitilde}[1]{\tilde{\phi}(#1)}
\newcommand{\tr}{\mathrm{tr}}
\newcommand{\tar}{\triangleright}

\begin{document}
\renewcommand{\thefootnote}{\fnsymbol{footnote}}
\begin{titlepage}
\vspace*{0mm}
\hfill
{\hfill \begin{flushright} 
YITP-07-14 
\end{flushright}
  }

\vspace*{10mm}

\begin{center}
{\LARGE {\LARGE  
Braided quantum field theories \\ and their symmetries}}
\vspace*{15mm}

 {\large Yuya Sasai}
\footnote{ e-mail: sasai@yukawa.kyoto-u.ac.jp}
 {\large and~ Naoki Sasakura}
\footnote{ e-mail: sasakura@yukawa.kyoto-u.ac.jp}

\vspace*{7mm}
{\large {\it 
Yukawa Institute for Theoretical Physics, Kyoto University, \\ 
 Kyoto 606-8502, Japan 
}} \\

\end{center}

\vspace*{1.3cm}

\begin{abstract}
Braided quantum field theories proposed by Oeckl can provide a framework
for defining quantum field theories having Hopf algebra symmetries.
In quantum field theories, symmetries lead to non-perturbative relations among 
correlation functions. 
We discuss Hopf algebra symmetries and 
such relations in braided quantum field theories.
We give the four algebraic conditions between Hopf algebra symmetries and 
braided quantum field theories, which are required for the relations to hold. As 
concrete examples, we apply our discussions 
to the Poincar\'e symmetries of two examples of noncommutative field theories. 
One is the effective quantum field theory of three-dimensional quantum gravity coupled with spinless particles given by Freidel and Livine, and the other is noncommutative field theory on Moyal plane.
We also comment on quantum field theory on {$\kappa$}-Minkowski spacetime.
\end{abstract}

\end{titlepage}
\newpage

\section{Introduction}
\renewcommand{\thefootnote}{\arabic{footnote}}
\setcounter{footnote}{0}
Symmetry is one of the most important notions in quantum field theory.
In many examples, 
it is useful in investigating properties of quantum field theories non-perturbatively,
is a guiding principle in constructing field theories for various purposes such
as grand unification, 
or gives powerful methods in finding exact solutions.
It also plays important roles in actual renormalization procedures. 
Therefore it should be interesting to study symmetries also in noncommutative 
field theories 
\cite{Snyder:1946qz,Yang:1947ud,Connes:1990qp,Doplicher:1994tu,Seiberg:1999vs}, 
which may result from some quantum gravity effects \cite{Garay:1994en}.

A difficulty in the study in this direction is the apparent violation 
of basic symmetries such as Poincar\'e symmetry in the noncommutativity of spacetime.
For example, the Moyal plane $[x^\mu,x^\nu]=i\theta^{\mu\nu}$ is translational invariant, 
but is not Lorentz or rotational invariant.
Another example is the three-dimensional spacetime with
noncommutativity $[x^i,x^j]=i\kappa \epsilon^{ijk}x_k\ (i,j,k=1,2,3)$ 
\cite{Sasakura:2000vc,Madore:2000en,Freidel:2005ec,Imai:2000kq} with a
noncommutativity parameter $\kappa$.
This noncommutative spacetime is Lorentz-invariant, but is not invariant under the 
translational transformation $x^i\to x^i+a^i$ with $c$-number $a^i$.  
In fact, a naive construction of noncommutative quantum field theory on this spacetime
leads to rather disastrous violations of energy-momentum conservation \cite{Imai:2000kq}: 
the violations coming from the non-planar diagrams do not vanish in the 
commutative limit $\kappa\rightarrow 0$ as in the UV/IR mixing phenomena
\cite{Minwalla:1999px}. 

In recent years, however, there has been interesting conceptual progress in
understanding symmetries in noncommutative field theories:
the symmetry transformations in noncommutative spacetime are not the usual 
Lie-algebraic type, but should be generalized to have Hopf algebraic structures.  
The Moyal plane was pointed out to be invariant under the twisted Poincar\'e transformation in
\cite{Chaichian:2004za,Wess:2003da,Koch:2004ud} and under the twisted diffeomorphism in \cite{Aschieri:2005yw,Aschieri:2005zs,Calmet:2005qm,Kobakhidze:2006kb}. There have been various proposals to implement the twisted Poincar\'e invariance in quantum field theories 
\cite{Chaichian:2004yh,Chaichian:2005yp,Balachandran:2005eb,Balachandran:2005pn,Lizzi:2006xi,
Tureanu:2006pb,Zahn:2006wt,Bu:2006ha,Abe:2006ig,Balachandran:2006pi,Fiore:2007vg,
Joung:2007qv}. 
As for the noncommutative spacetime with
$[x^i,x^j]=i\kappa \epsilon^{ijk}x_k$, a noncommutative quantum field theory was
derived as the effective field theory of three-dimensional quantum gravity 
with matters \cite{Freidel:2005bb}. 
Its essential difference from the naive construction mentioned above is the nontrivial braiding for each crossing in non-planar Feynman diagrams.
With this braiding, there exists a kind of conserved energy-momentum in the amplitudes,
and the energy-momentum operators have Hopf algebraic structures. 
 
Our aim of this paper is to systematically understand these Hopf algebraic symmetries and 
their consequences in noncommutative field theories in the framework of braided quantum
field theories proposed by Oeckl \cite{Oeckl:1999zu}. In the usual quantum field theories,
symmetries give non-perturbative relations among correlation functions. We will see that such relations have natural extensions 
to the Hopf algebraic symmetries in braided quantum field theories, 
and will obtain the four
conditions for the relations to hold. These conditions should be interpreted 
as the criteria of the symmetries in braided quantum field theories.

This paper is organized as follows. 
In the following section, we review braided quantum field theory.
This review part follows faithfully the original paper \cite{Oeckl:1999zu}, but figures are more extensively
used in the proofs and the explanations to make this paper self-contained and intuitively 
understandable.
We start with braided category and braided Hopf algebra. Then correlation functions of braided quantum field theory are represented in terms of them. Finally braided Feynman rules are given.

In Section \ref{sec:sym}, we first review the axioms of \textit{action}\footnote{We use the italic symbol to distinguish it from the action $S$.} of an algebra on vector spaces. Then we consider the relations among correlation functions in braided quantum field theory. We find that four algebraic conditions are required for the relations to hold.
Then, as concrete examples, we discuss whether the noncommutative field theories 
mentioned above have the Poincar\'e symmetry by checking the four conditions. 
In the former case, we find that the twisted Poincar\'{e} symmetry is implemented 
only after the introduction of a non-trivial braiding factor, which agrees with
the previous proposal in \cite{Balachandran:2005eb,Oeckl:2000eg}.
In the latter case, we find that the theory has a kind of translational symmetry, 
which is different from the usual one by multi-field contributions. 
We also give some examples of such relations among correlation functions and the implications.

The final section is devoted to summary and comments. We comment on quantum field theory on $\kappa$-Minkowski spacetime whose noncommutativity of coordinates is $[x^0,x^j]=\frac{i}{\kappa}x^j \ (j=1,2,3)$ \cite{Freidel:2006gc}. 

\section{Review of braided quantum field theory} \label{sec:braidedqft}
\subsection{Braided categories and braided Hopf algebras} \label{sec:braidedcat}
First of all, we review braided categories and braided Hopf algebras 
\cite{Oeckl:1999zu,Majid:1996kd}. Braided categories are composed of an object $X$, which is a vector space, a dual object $X^{\ast}$, which is a dual vector space, and morphisms
\begin{align}
\ev ~~&:~~ X^{\ast} \otimes X \to \Bbbk  ~~~\text{(evaluation)},\\
\coev ~~&:~~ \Bbbk  \to X\otimes X^{\ast}    ~~~\text{(coevaluation)},
\label{eq:defevcoev}
\end{align}
where $\Bbbk$ is a c-number. The composition of the two morphisms in an obvious way makes the identity. Then the braided categories have also an invertible morphism
\begin{equation}
\psi_{V,W}~~:~~V\otimes W \to W\otimes V ~~~\text{(braiding)},
\label{eq:defbraid}
\end{equation}
where $V,W$ are any pair of vector spaces. Generally the inverse of braiding is not equal to the braiding itself.

The braiding is required to be compatible with the tensor product such that
\begin{align}
\psi_{U,V\otimes W}&=(\mathrm{id} \otimes \psi_{U,W})\circ (\psi_{U,V}\otimes \mathrm{id}), \nonumber \\
\psi_{U\otimes V,W}&=(\psi_{U,W}\otimes \mathrm{id})\circ (\mathrm{id} \otimes \psi_{V,W}). \label{eq:axiomofbraid}
\end{align}
Then the braiding is also required to be \textit{intersectional} under any morphisms
in a Hopf algebra. For example, 
\begin{align}
\psi_{Z,W}(Q\otimes \id)&=(\id \otimes Q)\psi_{V,W} ~~\text{for any} ~Q:V\to Z,  \notag \\ 
\psi_{V,Z}(\id\otimes Q)&=(Q \otimes \id)\psi_{V,W} ~~\text{for any} ~Q:W\to Z, \label{eq:commuteaxm}
\end{align}
where $Z$ is a vector space.

\begin{figure}
\begin{center}
\includegraphics[scale=0.7]{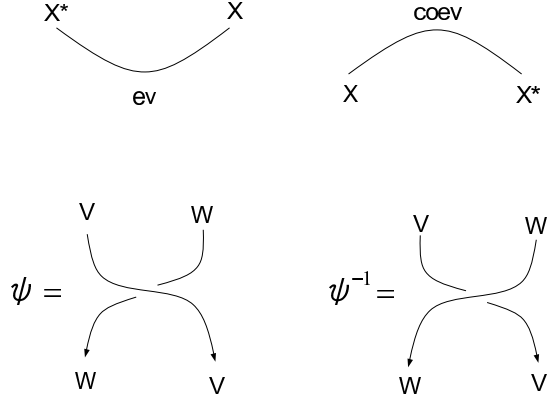}
\end{center}
\caption{The evaluation, coevaluation, braiding and its inverse.}
\label{fig:morph}
\end{figure}

\begin{figure}
\begin{center}
\includegraphics[scale=0.7]{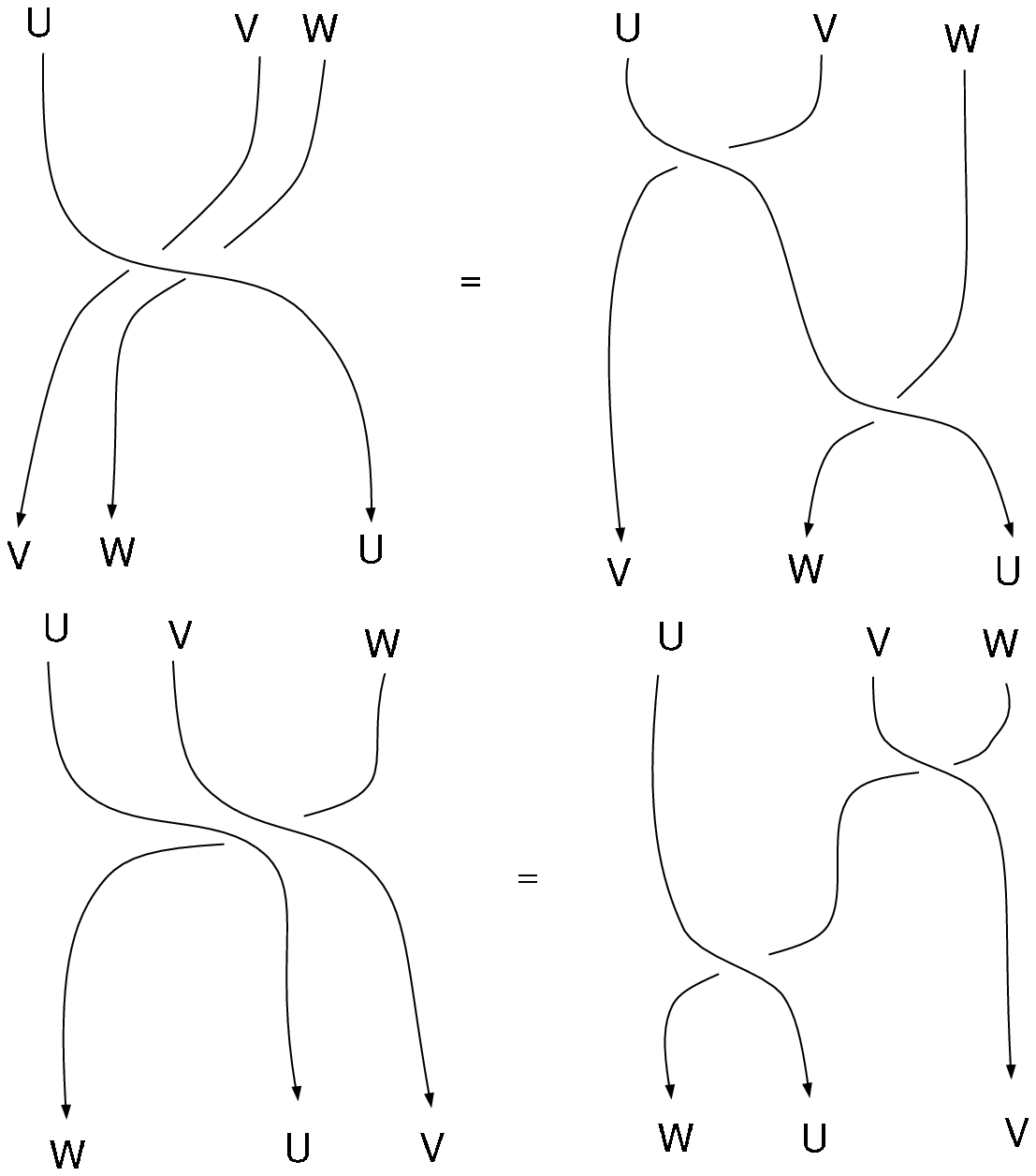}
\end{center}
\caption{The axioms of braiding (\ref{eq:axiomofbraid}).}
\label{fig:braidaxm}
\end{figure}

\begin{figure}
\begin{center}
\includegraphics[scale=0.6]{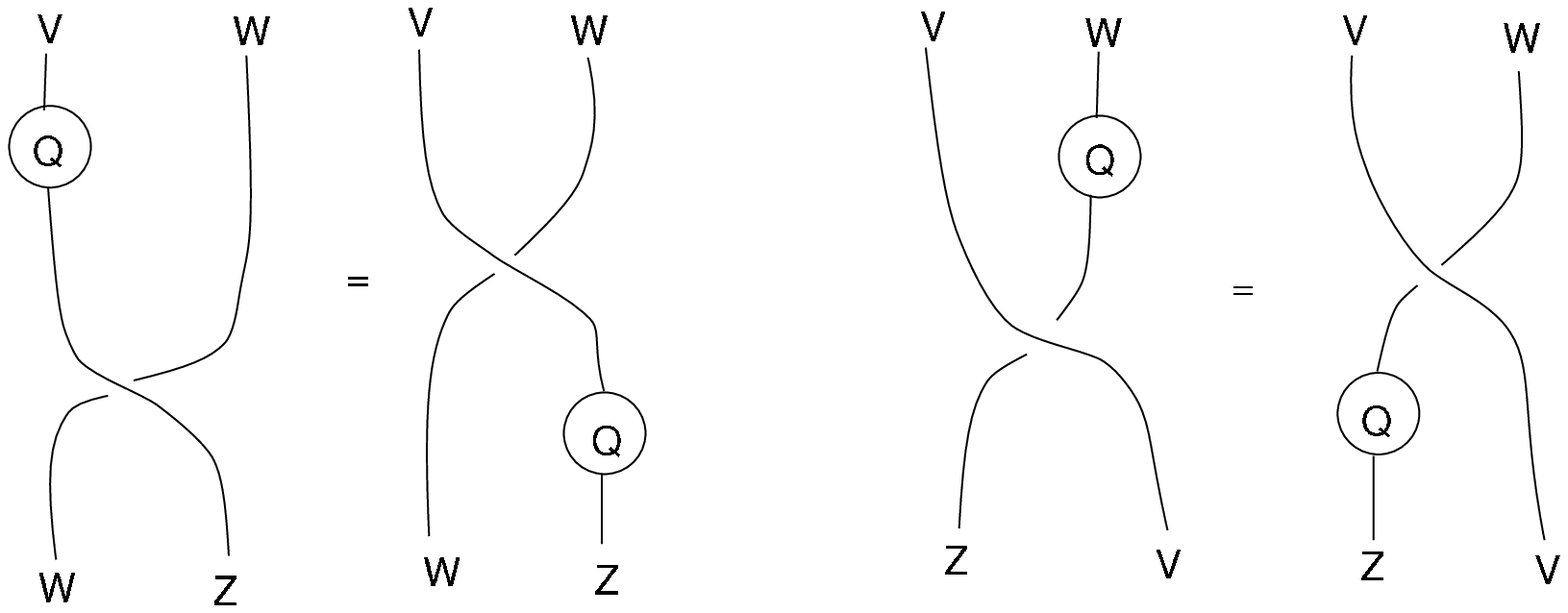}
\end{center}
\caption{The axioms of braiding (\ref{eq:commuteaxm}).}
\label{fig:commuteaxm}
\end{figure}

We can represent these axioms in pictorial ways \cite{Majid:1992sn}. We write the morphisms, ev, coev, $\psi$, downwards as in Figure \ref{fig:morph}. 
Thus the axioms (\ref{eq:axiomofbraid}) are represented as in Figure \ref{fig:braidaxm}, and the axioms (\ref{eq:commuteaxm}) are represented as in Figure \ref{fig:commuteaxm}.

\begin{figure}
\begin{center}
\includegraphics[scale=0.6]{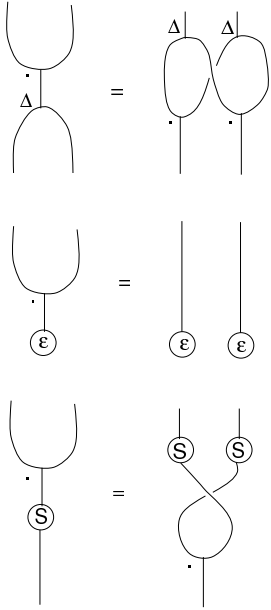}
\end{center}
\caption{The axioms of coproduct, counit, antipode for products.}
\label{fig:prodaxm}
\end{figure}

Next we consider the polynomials of $X$,
\begin{equation}
\Xh :=\bigoplus_{n=0}^{\infty}X^n,~~\text{with}~~X^0:=\mathbf{1}~~\text{and}~~X^n:=\underbrace{X\otimes \cdots \otimes X}_{n ~\text{times}},
\end{equation}
where $\mathbf{1}$ is the trivial one-dimensional space. $\Xh$ naturally has the structure of a braided Hopf algebra via
\begin{align}
\cdot ~~(\text{product})&: \Xh \otimesh \Xh \to \Xh , \\
\eta ~~(\text{unit})&: \Bbbk \to \Xh ~;~ \eta(1)=\mathbf{1}, \\
\Delta ~~(\text{coproduct})&: \Xh \to \Xh \otimesh \Xh ~;~ \Delta \phi =\phi \otimesh \mathbf{1} +\mathbf{1} \otimesh \phi, ~~\text{and} ~\Delta(\mathbf{1})=\mathbf{1}\otimesh \mathbf{1}, \label{eq:copro} \\
\epsilon ~~(\text{counit})&: \Xh \to \Bbbk ~;~ \epsilon (\phi)=0, ~~\text{and} ~\epsilon (\mathbf{1})=1, \\
S ~~(\text{antipode})&:\Xh \to \Xh ~;~ S \phi =-\phi, ~~\text{and} ~S (\mathbf{1})=\mathbf{1},
\end{align}
where $\phi \in X$. The tensor product $\otimes$ is the same as the usual product of $X$s,
while the new tensor product $\otimesh $ is the tensor product of $\Xh$s. The coproduct $\Delta$, counit $\epsilon$, antipode $S$  of the products of $X$s are defined inductively by
\begin{align}
\Delta \circ \cdot &= (\cdot \otimesh \cdot )\circ (\id \otimesh \psi \otimesh \id)\circ (\Delta \otimesh \Delta ), \label{eq:coproaxm} \\
\epsilon \circ \cdot &=\cdot \circ (\epsilon \otimesh \epsilon), \\
S\circ \cdot &=\cdot \circ \psi \circ (S \otimesh S).
\end{align}
These axioms are diagrammatically represented in Figure \ref{fig:prodaxm}.

\subsection{Braided quantum field theory}
Next we represent braided quantum field theory \cite{Oeckl:1999zu} in terms of the braided category and the braided Hopf algebra. We take the vector space $X$ as the space of a field $\phi(x)$,
where $x$ denotes a general index for independent modes of the field. Thus $\Xh$ is the space of polynomials of the fields such as $\phi(x_1)\phi(x_2)\cdots \phi(x_n)$, 
and $\mathbf{1}$ correspond to the constant field of unit. We also take the dual vector space $\X$ as the space of differentials $\delta/\delta \phi(x)$. We take the evaluation and the coevaluation as follows,
\begin{align}
\ev &: \frac{\delta}{\delta \phi(x)} \otimes \phi(x')\to \delta(x-x'), \\
\coev &: 1 \to \int_x  ~\phi(x) \otimes \frac{\delta}{\delta \phi(x)},
\end{align}
where the distribution and the integration should symbolically be understood, and 
their detailed forms, which may contain non-trivial measures, depend on each case.

\begin{figure}
\begin{center}
\includegraphics[scale=0.5]{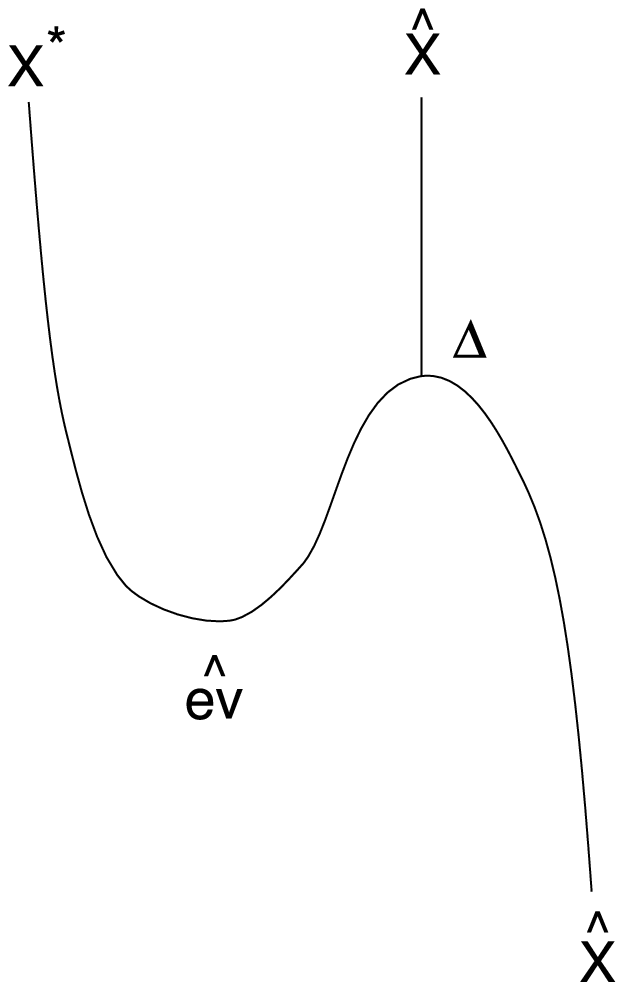}
\end{center}
\caption{The differentials on $\Xh$.}
\label{fig:diff}
\end{figure}

The differential on $\Xh$ is defined by 
\begin{equation}
\mathrm{diff}:= (\widehat{\ev}\otimes \id)\circ (\id \otimes \Delta); ~~\X\otimes \Xh \to \Xh, \label{eq:evhat}
\end{equation}
where 
\begin{equation}
\widehat{\ev}|_{\X \otimes X^n}=
\begin{cases}
&\!\!\! \ev ~~\text{for} ~n=1, \\
&\!\! 0 ~~~\text{for} ~n\neq 1.
\end{cases}
\end{equation}
Diagrammatically this is given by Figure \ref{fig:diff}.

To see whether the map $\diff$ gives really the differential of products, let us compute the differential of $\phi(x)\phi(y)$ as a simple example 
using the definition (\ref{eq:evhat}).
This becomes
\begin{align}
\diff \bigg(\dd{x'} \otimes \phi(x)\phi(y)\bigg)&=(\widehat{\ev}\otimes \id)\circ (\id \otimes \Delta)\bigg(\dd{x'} \otimes \phi(x)\phi(y)\bigg) \notag \\
&=(\widehat{\ev}\otimes \id)\circ \bigg(\dd{x'} \otimes \Delta(\phi(x)\phi(y))\bigg) \notag \\
&=(\widehat{\ev}\otimes \id)\circ \bigg(\dd{x'} \otimes (\phi(x)\phi(y)\otimesh \1\notag \\
&~~~~~+\phi(x)\otimesh \phi(y)+\psi (\phi(x)\otimesh \phi(y))+\1\otimesh \phi(x)\phi(y))\bigg) \notag \\
&=\delta(x'-x)\otimes \phi(y)+(\widehat{\ev}\otimes \id)\circ \bigg(\dd{x'}\otimes \psi (\phi(x)\otimesh \phi(y))\bigg),
\end{align}
where we have used the axiom (\ref{eq:coproaxm}) in deriving the third line. 
If the braiding is trivial, we find that the differential (\ref{eq:evhat}) satisfies 
the usual Leibniz rule.

\begin{figure}
\begin{center}
\includegraphics[scale=0.6]{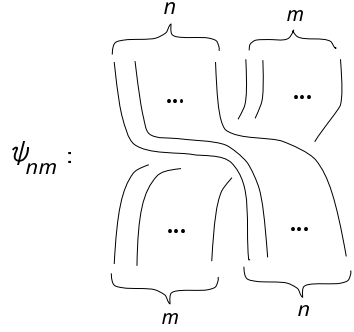}
\end{center}
\caption{Diagram of $\psi_{n,m}$.}
\label{fig:psinm}
\end{figure}

Generally we find a braided Leibniz rule
\begin{equation}
\partial (\alpha \beta)=\partial (\alpha) \beta +\psi^{-1}(\partial \otimes \alpha )(\beta ) \label{eq:BLR1}
\end{equation}
and
\begin{equation}
\partial (\alpha )=(\ev \otimes \id^{n-1})(\partial \otimes [n]_{\psi}\alpha ), \label{eq:BLR2}
\end{equation}
where $\partial \in \X$, $\alpha,\beta \in \Xh $, and we have used 
a simplified notation
\begin{equation}
\partial (\alpha):=\diff (\partial \otimes \alpha).
\end{equation} 
Here $n$ is the degree of $\alpha$, and $[n]_{\psi}$ is called a braided integer defined
by
\begin{equation}
[n]_{\psi}:=\id^n+\psi \otimes \id^{n-2}+\cdots +\psi_{n-2,1}\otimes \id+\psi_{n-1,1},
\end{equation}
where $\psi_{n,m}$ is a braiding morphism given in Figure \ref{fig:psinm}.

 The proofs of the formula (\ref{eq:BLR1}), (\ref{eq:BLR2}) are in Appendix \ref{sec:app1}.

Now we define a Gaussian integration, which defines the path integral. 
The definition is given by
\begin{equation}
\int \partial (\alpha w):=0 ~~\text{for}~~\partial \in \X, \alpha \in \Xh, \label{eq:int}
\end{equation}
where $w\in \Xh$ is a Gaussian weight. In field theory, $w$ is the exponential of the free part of the action, $e^{-S_0}$.

In order to obtain a formula for correlation functions, 
we define a morphism $\gamma : \X\to X$ such that 
\begin{equation}
\partial (w):=-\gamma(\partial )w. \label{eq:gamma}
\end{equation}
This morphism is assumed to be commutative with the braiding as in (\ref{eq:commuteaxm}).
If $w=e^{-S_0}$, $\gamma(\partial)=\partial (S_0)$. In field theory, 
this is the kinetic part of the action, or the inverse of the propagator.  

Starting from (\ref{eq:int}), we can represent correlation functions of a free field theory
in terms of the braided category and the braided Hopf algebra. This is the analog of the Wick theorem in 
braided quantum field theory. 
The definition of the free $n$ point correlation function is given by
\begin{equation}
Z^{(0)}_n(\alpha):=\frac{\int \alpha w}{\int w},
\end{equation}
where the degree of $\alpha $ is $n$. Algebraically, this is given by
\begin{align}
\Zfreed{2}&= \ev\circ (\gamma^{-1} \otimes \id) \circ \psi \label{eq:bwick1}, \\
\Zfreed{2n}&=(\Zfreed{2})^n\circ [2n-1]^{'}_{\psi}!! \label{eq:bwick2}, \\
\Zfreed{2n-1}&=0 , \label{eq:bwick3}
\end{align} 
where 
\begin{align}
[2n-1]^{'}_{\psi}!!&:=([1]'_{\psi}\otimes \id^{2n-1})\circ ([3]'_{\psi}\otimes \id^{2n-3})\circ \cdots \circ ([2n-1]'_{\psi}\otimes \id), \\
[n]^{'}_{\psi}&:=\id^n+\id^{n-2}\otimes \psi^{-1}+\cdots +\psi^{-1}_{1,n-1} \notag \\
&=\psi_{1,n-1}^{-1}\circ [n]_{\psi}.
\end{align}
The proofs of (\ref{eq:bwick1}), (\ref{eq:bwick2}), (\ref{eq:bwick3}) are in Appendix \ref{sec:app2}.

Next we consider correlation functions with the existence of an interaction. For $S=S_0+\lambda S_{int}$, a correlation function is perturbatively given by
\begin{align}
Z_n(\alpha )&=\frac{\int \alpha e^{-S}}{\int e^{-S}} \notag \\
&=\frac{\int \alpha (1-\lambda S_{int}+\cdots ) e^{-S_0}}{\int (1-\lambda S_{int}+\cdots )e^{-S_0}}, \label{eq:corint}
\end{align}
where $\alpha \in X^{n}$. Introducing a morphism $S_{int} : \Bbbk \to X^{k} $, where $k$ is the degree of $S_{int}$, the correlation function is algebraically given by
\begin{align}
Z_n=\frac{Z_n^{(0)}-\lambda Z_{n+k}^{(0)} \circ (\id^n \otimes S_{int})+\frac{1}{2}\lambda^2 Z_{n+2k}^{(0)} \circ (\id^n \otimes S_{int}\otimes S_{int})+\cdots }{1-\lambda Z_k^{(0)} \circ S_{int}+\frac{1}{2}\lambda^2 Z_{2k}^{(0)} \circ (S_{int}\otimes S_{int})+\cdots }. \label{eq:fullcorrelation}
\end{align}
Acting $Z_n$ on $\alpha \in X^n$, we obtain the correlation function (\ref{eq:corint}). 
One can obviously extend $S_{int}$ to include various interaction terms.

\subsection{Braided Feynman rules}
\begin{figure}
\begin{center}
\includegraphics[scale=0.5]{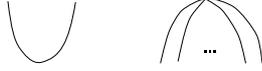}
\end{center}
\caption{Propagator (left) and vertex (right).}
\label{fig:propver}
\end{figure}

\begin{figure}
\begin{center}
\includegraphics[scale=0.5]{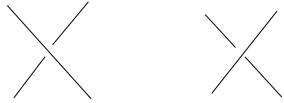}
\end{center}
\caption{The braiding $\psi$ (left) and its inverse $\psi^{-1}$ (right).}
\label{fig:braiddgm}
\end{figure}

\begin{figure}
\begin{center}
\includegraphics[scale=0.3]{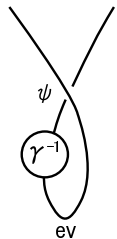}
\end{center}
\caption{The propagator, which is abbreviated in the left figure of Figure \ref{fig:propver}.}
\label{fig:prop}
\end{figure}

From the results in the preceding subsection, 
a correlation function can be represented by summation of diagrams 
obeying the following rules below.
\begin{itemize}
	\item 
An $n$-point function $Z_n$ is a morphism $X^{n} \to \Bbbk$. 
Thus a Feynman diagram starts with $n$ strands at the top and must be closed at the bottom. 
	\item The propagator $Z^{(0)}_2: X\otimes X \to \Bbbk$ is represented by the left of Figure \ref{fig:propver}, which is the abbreviation of Figure \ref{fig:prop}.
	\item The interaction  vertex $S_{int}: \Bbbk \to X^{k} $ is represented by the right of Figure \ref{fig:propver}. Generally the order of the strands is noncommutative.
	\item The two kinds of crossings, which are represented in Figure \ref{fig:braiddgm}, correspond to the braiding and its inverse.
	\item Any Feynman diagram is built out of propagators, vertices, and crossings, and is closed at the bottom.
	
\end{itemize}


\section{Symmetries in braided quantum field theory} \label{sec:sym}
In this section, 
we discuss symmetries in braided quantum field theory. In order to represent symmetry
transformations on fields, 
we review general description of an \textit{action} in Section \ref{subsec:action}. In Section \ref{subsec:wtid}, we study relations among correlation functions. 
We find four conditions for such relations to follow from a symmetry algebra.
In Section \ref{subsec:3dgravity} and \ref{subsec:twistednc}, we treat two examples of (braided) noncommutative field theories and discuss their Poincar\'e symmetries.

\subsection{General description of an \textit{action}} \label{subsec:action}
We review an \textit{action} of a general Hopf algebra on vector spaces in a mathematical language \cite{Majid:1996kd,Klimyk:1997eb}. 

An \textit{action} $\alpha_V$ is a map $\alpha_V:\mathcal{A}\otimes V\to V$, where $\mathcal{A}$ is an arbitrary Hopf algebra and $V$ is a vector space (in our case, $\mathcal{A}$ is a symmetry algebra, and $V=X$ or $\X$). We will denote the coproduct and the counit of the Hopf algebra\footnote{We omit the antipode.} by $\Delta'$ and $\epsilon'$ to distinguish them from those of 
the braided Hopf algebra of fields in Section \ref{sec:braidedqft}.
We do not write all the axioms of an \textit{action}, 
but our important axioms are the following.
\begin{itemize}
    \item $\alpha_V$ satisfies the following condition.
     \begin{equation}
     \alpha_V \circ (\cdot \otimes \id)=\alpha_V \circ (\id \otimes \alpha_V ),  \label{eq:transientofalg}
     \end{equation}
     where the equality acts on $\A\otimes \A \otimes V$. This means that $\alpha_V ((a\cdot b) \otimes V)=\alpha_V (a \otimes (\alpha_V(b \otimes V)))$, where $a, b\in \A$. In short we can write this as
     \begin{equation}
     (a \cdot b)\tar V=a\tar (b \tar V).
     \end{equation}
     
     \item An \textit{action} on $\1$, which is in a vector space, is defined by
     \begin{equation}
     \alpha_V(a \otimes \1)=\epsilon'(a)\1, \label{eq:unitaction}
     \end{equation}
     where $\epsilon'(a)$ is the counit of an algebra $a\in \A$.
	\item An \textit{action} on a tensor product of vector spaces $V,W$ is defined by
\begin{equation}
\alpha_{V\otimes W}(a):=((\alpha_V\otimes \alpha_W)\circ \Delta')(a)=\sum_i \alpha_V(a^i_{(1)})\otimes \alpha_W(a^i_{(2)}), ~a\in \mathcal{A},
\end{equation}
where $\Delta' (a)=\sum_i a^i_{(1)}\otimes a^i_{(2)}$ is the coproduct of the Hopf algebra $\mathcal{A}$.  In the case of a usual Lie-algebraic transformation, its coproduct 
is given by $\Delta' (a)=a \otimes \1+\1\otimes a$, where $\1$ is in $\mathcal{A}$. This gives the usual Leibnitz rule.
    \item Since a Hopf algebra has the coassociativity that
    \begin{equation}
    ((\Delta' \otimes \id)\circ \Delta')(a)=((\id \otimes \Delta')\circ \Delta')(a),
    \end{equation}
    the \textit{action} on a tensor product of vector spaces, which is obtained by
the multiple operations of $\Delta'$ on $a$, is actually unique. 
An important consequence is that one 
    can divide the \textit{action} on a tensor product of vector spaces as
     \begin{align}
     a\tar &(V_1\otimes \cdots \otimes V_{k-1}\otimes V_{k}\otimes \cdots \otimes V_{n})= \notag \\
     &\sum_i a_{(1)}^i\tar (V_1\otimes \cdots \otimes V_{k-1})\otimes a_{(2)}^i\tar (V_{k}\otimes \cdots \otimes V_{n}) \label{eq:algasso}
     \end{align}
     for any $k$.

\end{itemize}

\subsection{Symmetry relations among correlation functions and their algebraic descriptions} \label{subsec:wtid}
The expression of the correlation functions (\ref{eq:fullcorrelation}) is perturbative 
in interactions, but is a full order algebraic description. Therefore we can discuss the 
symmetry of the theory and the implied relations among correlation functions 
by using this expression.
We may even expect that the relations will hold non-perturbatively.
  
In usual quantum field theory, if a field theory has a certain symmetry, there is 
a relation among the correlation functions in the form,
\begin{equation}
\sum_{i=1}^n \langle \phi(x_1)\cdots \delta_a \phi(x_i)\cdots \phi(x_n)\rangle=0, \label{eq:wtgeneral}
\end{equation}
where $\delta_a \phi(x)$ is a variation of a field under a transformation $a$, on the 
assumption that the path integral measure and the action are invariant under the transformation.

\begin{figure}
\begin{center}
\includegraphics[scale=0.6]{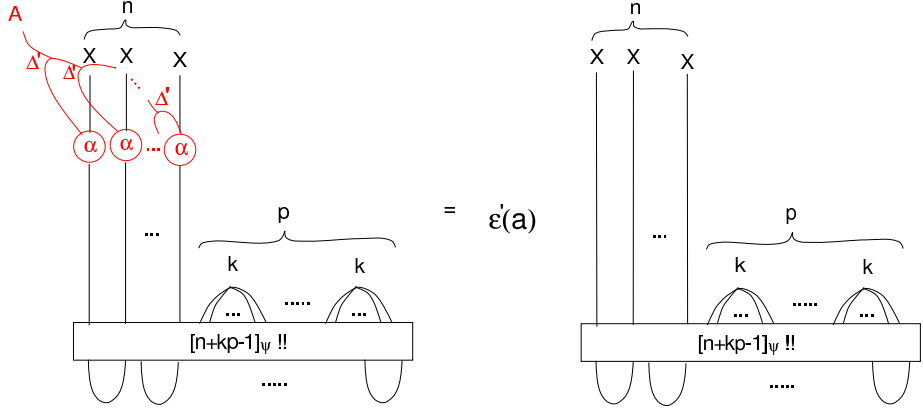}
\end{center}
\caption{A relation among correlation functions in the diagrammatic representation. $n$ is the number of external legs, $k$ is the order of the interaction, and $p$ is the order of the perturbation. $n+kp$ is even.}
\label{fig:wtid}
\end{figure}

If the coproduct of a symmetry algebra is not the usual Lie-algebraic type
and thus the Leibniz rule is deformed, the relation will generally have the form,
\begin{align}
&c_a^{(bi)}\langle \phi(x_1)\cdots \delta_b \phi(x_i)\cdots \phi(x_n)\rangle \notag \\
 +& c_a^{(bi)(cj)}\langle \phi(x_1) \cdots \delta_b \phi(x_i) \cdots \delta_c\phi(x_j) \cdots \phi(x_n)\rangle \notag \\
+&  c_a^{(bi)(cj)(dk)}\langle \phi(x_1) \cdots \delta_b\phi(x_i) \cdots \delta_c\phi(x_j)\cdots \delta_d\phi(x_k) \cdots \phi(x_n)\rangle \notag \\
+&\cdots =0, \label{eq:wtidgeneral}
\end{align}
where $c_a^{\cdots}$ are some coefficients. Its essential difference from (\ref{eq:wtgeneral}) is the multi-field contributions.
In our algebraic language, the relation can be written as
\begin{equation}
Z_n(a\tar \chi)=\epsilon'(a) Z_n(\chi),~~~~\mathrm{for} ~a\in \A, ~\chi \in X^n.
\label{eq:wtbraid}
\end{equation}
This is  equivalent to Figure \ref{fig:wtid} in our diagrammatic representation. Then we consider what an algebraic structure is required for (\ref{eq:wtbraid}) to hold
for any $a$ and $\chi$, i.e. the theory is invariant under the Hopf algebra transformation
$\A$. 

Let us write the coproduct of an element $a\in\mathcal{A}$ as 
\begin{equation}
\Delta' (a)=\sum_s f^s \otimes g^s,  \label{eq:generalcopro}
\end{equation}
where $f^s,g^s\in \A$.
Since the coproduct must satisfy the Hopf algebra axiom \cite{Majid:1996kd}, 
\begin{equation}
(\epsilon' \otimes \id )\Delta'( a)=(\id \otimes \epsilon' )\Delta' (a)=a, 
\label{eq:axiomcounit}
\end{equation}
$f^s, g^s$ must satisfy
\begin{align}
\sum_s \epsilon' (f^s)\otimes g^s=\sum_s f^s \otimes \epsilon'(g^s)=a. \label{eq:fgcond}
\end{align}

\begin{figure}
\begin{center}
\includegraphics[scale=0.6]{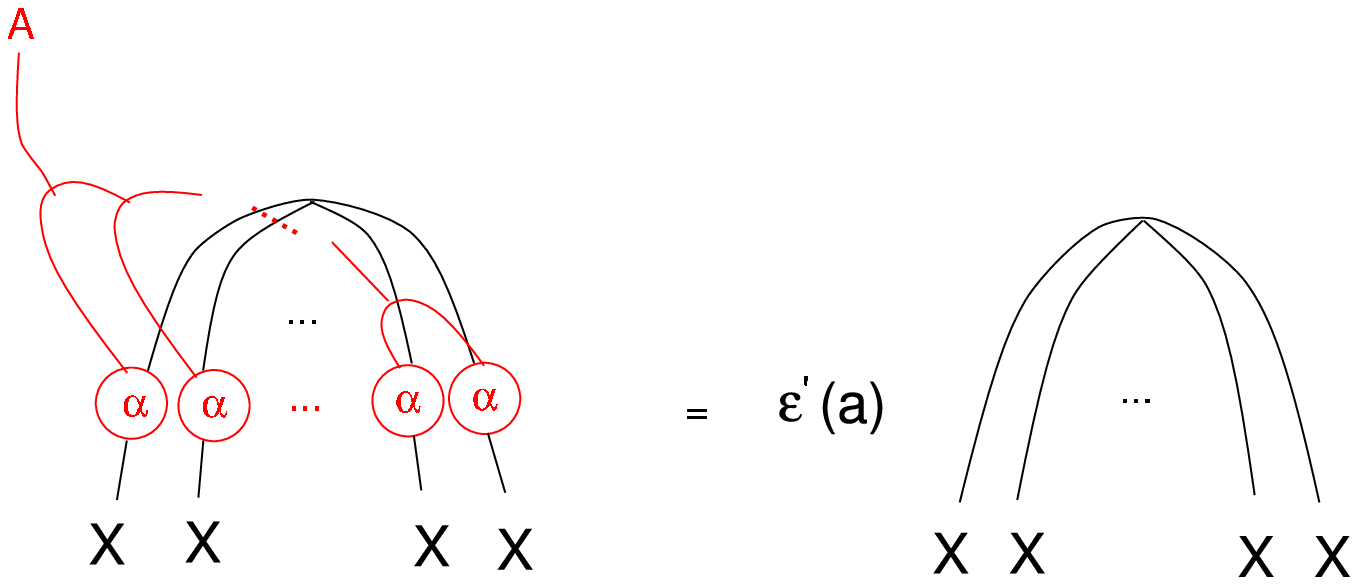}

\includegraphics[scale=0.6]{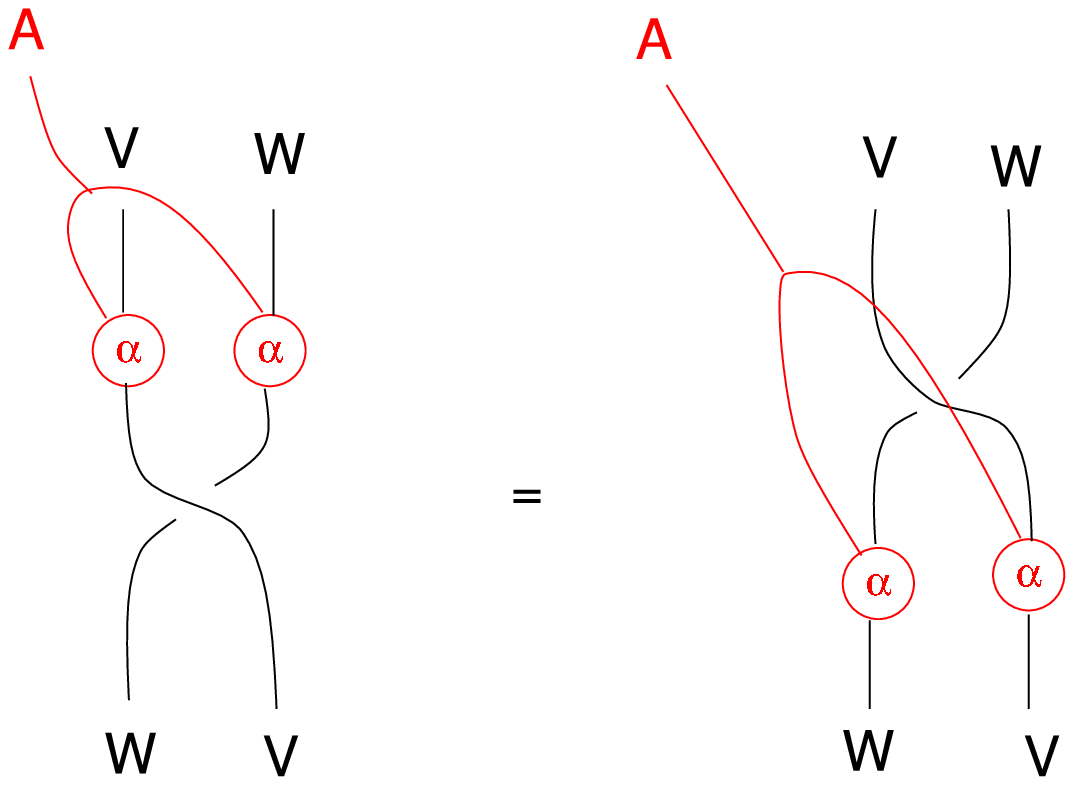}

\includegraphics[scale=0.6]{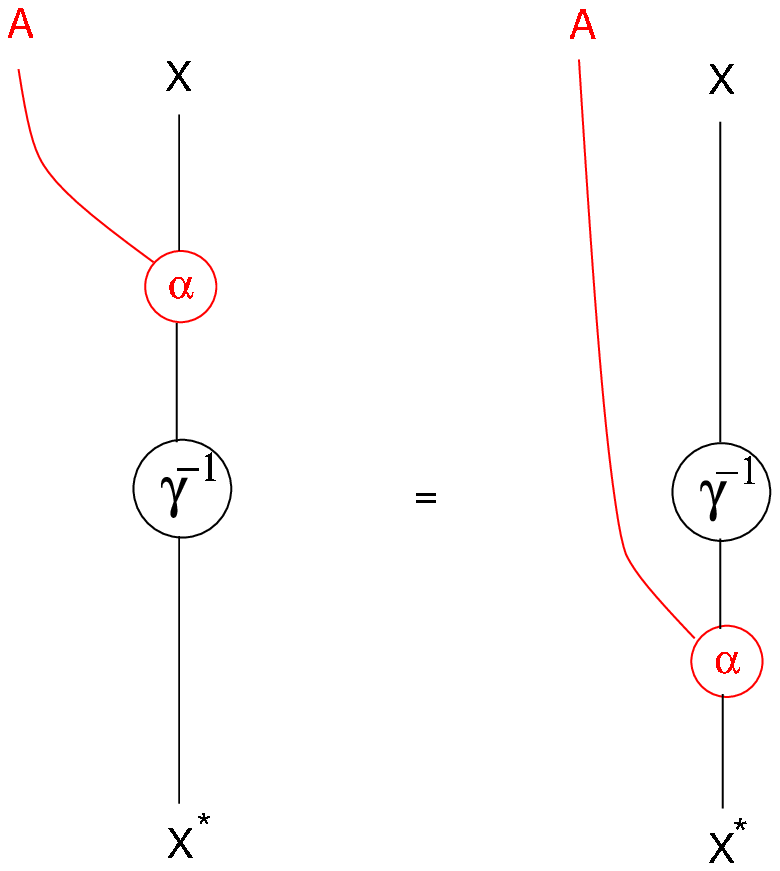}

\includegraphics[scale=0.6]{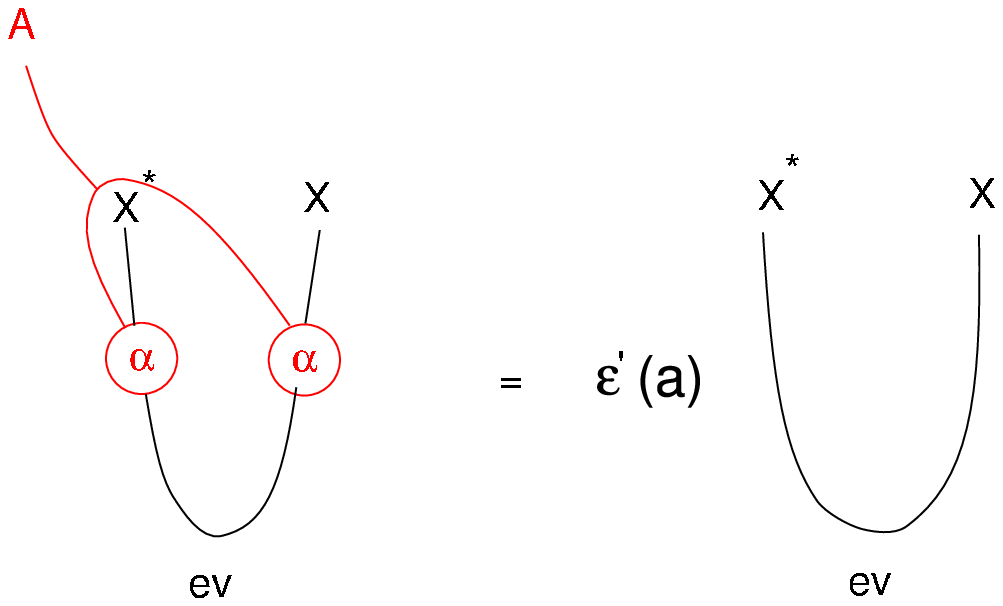}
\end{center}
\caption{Conditions 1,2,3, and 4.}
\label{fig:condition}
\end{figure}

For all the relations among correlation functions to hold, we find the following four conditions for any \textit{action} $a\in \A$. 
\begin{itemize}
	\item (Condition 1) $S_{int}$ must satisfy
	\begin{equation}
	a\tar S_{int}=\epsilon'(a) S_{int}. \label{eq:cond1}
	\end{equation}
	\item (Condition 2) The braiding $\psi$ is an intertwining operator. That is
	\begin{equation}
	\psi (a\tar (V\otimes W))=a \tar \psi(V\otimes W). \label{eq:cond2}
	\end{equation}
	\item (Condition 3) $\gamma^{-1}$ and $a$ are commutative,
	\begin{equation}
	a \tar (\gamma^{-1}(V))=\gamma^{-1}(a\tar V). \label{eq:cond3}
	\end{equation}
	\item (Condition 4) Under an \textit{action} $a$, the evaluation map follows
	\begin{equation}
	\ev (a\tar (\X\otimes X))=\epsilon'(a)\ev(\X\otimes X). \label{eq:cond4}
	\end{equation}
\end{itemize}
Condition 1 to 4 are diagrammatically represented in Figure \ref{fig:condition}.
It is clear that, when the algebra $\A$ is generated from a finite number of its 
independent elements, it is enough for these generators to satisfy these conditions.

Condition 1 is the requirement of the symmetry at the classical level for the interaction. We can extend this condition to
\begin{equation}
(a \tar X^n) \otimes S_{int}^p=a \tar(X^n\otimes S_{int}^p). \label{eq:remma1stcond}
\end{equation}

The proof is the following. From a coproduct (\ref{eq:generalcopro}) and its coassociativity (\ref{eq:algasso}), the right hand side of (\ref{eq:remma1stcond}) is equal to
\begin{align}
\sum_s(f^s\tar (X^n\otimes S_{int}^{p-1}))\otimes g^s\tar S_{int} \label{eq:totyuu}
\end{align}
Since Condition 1 implies 
\begin{equation}
g^s \tar S_{int}=\epsilon'(g^s) S_{int},
\end{equation}
(\ref{eq:totyuu}) becomes
\begin{align}
&\sum_s(f^s\tar (X^n\otimes S_{int}^{p-1}))\otimes \epsilon'(g^s) S_{int} \notag \\
&=a\tar (X^n\otimes S_{int}^{p-1})\otimes S_{int},
\end{align}
where we have used (\ref{eq:fgcond}). 
Iterating this procedure, we obtain the left-hand side of (\ref{eq:remma1stcond}).

Condition 2,3,4 can also be extended to
\begin{align}
[n+kp-1]_{\psi}!!\circ (a\tar X^{n+kp})&=a\tar [n+kp-1]_{\psi}!!~X^{n+kp},  \label{eq:braidintertwiner} \\
(\gamma^{-1} \otimes \id)^{\frac{n+pk}{2}}\circ (a\tar X^{n+kp})&=a \tar (\gamma^{-1} \otimes \id )^{\frac{n+pk}{2}}X^{n+kp} \label{eq:gammaintertwiner}, \\
\ev^{\frac{n+pk}{2}}(a\tar (\X\otimes X)^{\frac{n+pk}{2}}) &=\epsilon'(a)~\ev^{\frac{n+pk}{2}}(\X\otimes X)^{\frac{n+pk}{2}}. \label{eq:extended4cond}
\end{align}

\begin{figure}
\begin{center}
\includegraphics[scale=0.65]{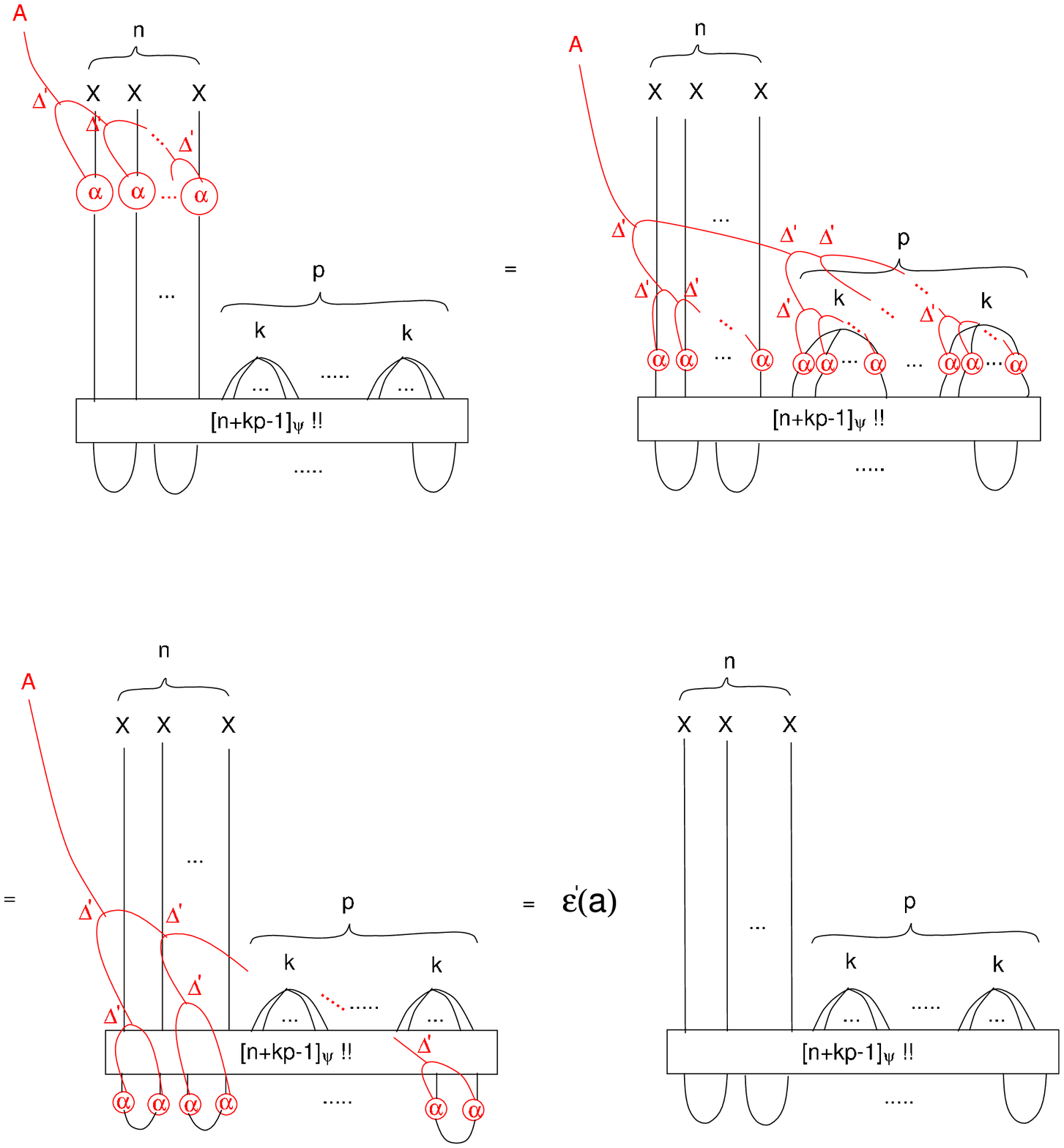}
\end{center}
\caption{A relation among correlation functions is satisfied if the four conditions (\ref{eq:cond1}), (\ref{eq:cond2}), (\ref{eq:cond3}), (\ref{eq:cond4}) are satisfied.}
\label{fig:wtid1}
\end{figure}

We can find that these extended conditions (\ref{eq:remma1stcond}), (\ref{eq:braidintertwiner}), (\ref{eq:gammaintertwiner}), (\ref{eq:extended4cond}) can be represented as in Figure \ref{fig:wtid1}. In the diagrammatic language, the relation among correlation functions 
holds if an \textit{action} can pass downwards through a Feynman diagram
 and satisfies (\ref{eq:unitaction}).

\subsection{Symmetries of the effective noncommutative field theory of three-dimensional
quantum gravity coupled with scalar particles} \label{subsec:3dgravity}
In this subsection, we discuss the Poincar\'e symmetry of the effective noncommutative field theory of three-dimensional quantum gravity coupled with scalar particles, which was obtained in \cite{Freidel:2005bb} by studying the Ponzano-Regge model \cite{Ponzano}
coupled with spinless particles. The symmetries of this theory is also known as $\mathrm{DSU}(2)$, which was discussed in \cite{Noui:2006kv,Noui:2006ku}. We first review the field theory 
\cite{Imai:2000kq,Freidel:2005bb}.

Let $\phi(x)$ be a scalar field on a three-dimensional space $x=(x^1,x^2,x^3)$. 
Its Fourier transformation is given by
\begin{equation}
\phi(x)=\int dg \tilde{\phi}(g) e^{\frac{i}{2\kappa }\mathrm{tr}(\mathcal{X}g)},
\end{equation}
where $\kappa$ is a constant, $\mathcal{X}=ix^i\sigma_i $, and $g=P^0-i\kappa P^i\sigma_i \in \mathrm{SO}(3)$\footnote{The identification $g\sim -g$ is implicitly assumed.} with
Pauli matrices $\sigma_i$. Here $\int dg$ is the Haar measure of $SO(3)$ and
$P^0=\pm\sqrt{1-\kappa^2P^iP_i}$ by definition. 
In the following discussions, we will only deal with the Euclidean case, but
the Lorentzian case can also be treated in a similar manner 
by replacing $\mathrm{SO}(3)$ with $\mathrm{SL}(2,R)$. 

The definition of the star product is given by
\begin{equation}
e^{\frac{i}{2\kappa }\tr(\mathcal{X}g_1)}\star e^{\frac{i}{2\kappa }\mathrm{tr}(\mathcal{X}g_2)}:=e^{\frac{i}{2\kappa }\mathrm{tr}(\mathcal{X}g_1g_2)}. \label{eq:defofstar}
\end{equation}
Differentiating both hands sides of (\ref{eq:defofstar}) with respect to $P_1^i:=P^i(g_1)$ and  $P_2^j:=P^j(g_2)$ and then taking the limit $P_1^i, P_2^i\to 0$, one finds 
the SO(3) Lie-algebraic space-time noncommutativity \cite{Sasakura:2000vc,Madore:2000en,Freidel:2005ec},
\begin{equation}
[x^i,x^j]_{\star}=2i\kappa\epsilon^{ijk}x_k. 
\label{eq:noncommutativity}
\end{equation}
For example, the action\footnote{Since in the Ponzano-Regge model the definition of the weight of partition function is $e^{iS}$ despite of Euclidean theory, the sign of the mass term is not the usual one. } of a $\phi^3$ theory is
\begin{equation}
S=\frac{1}{8\pi \kappa^3}\int d^3x\bigg[\frac{1}{2}(\partial_i\phi \star \partial_i\phi)(x)-\frac{1}{2}M^2(\phi \star \phi)(x)+\frac{\lambda}{3!}(\phi \star \phi \star \phi)(x) \bigg],
\end{equation}
where $M^2=\frac{\sin^2 m\kappa}{\kappa^2}$. Its momentum representation is 
\begin{align}
S=\frac{1}{2}\int dg \big(P^2(g)-&M^2\big)\phitilde{g}\phitilde{g^{-1}} \notag \\
&+\frac{\lambda }{3!}\int dg_1dg_2dg_3\delta(g_1g_2g_3)\phitilde{g_1}\phitilde{g_2}\phitilde{g_3},
\end{align}
from which it is straightforward to read the Feynman rules. 

Some quantum properties of this scalar field theory were analyzed 
in \cite{Imai:2000kq}. As can be seen from
(\ref{eq:noncommutativity}), 
the naive translational symmetry is violated. In fact, the violation is rather disastrous. 
There exists a kind of conserved energy-momentum in the amplitudes of 
the tree and the planar loop diagrams, but this energy-momentum is not conserved 
in the non-planar loop diagrams. Moreover, the violation of the energy-momentum
conservation
does not vanish in the commutative limit $\kappa\rightarrow 0$ due to 
a mechanism similar to the UV/IR phenomena \cite{Minwalla:1999px}.  
 
In the effective field theory of quantum gravity coupled with spinless particles,
however, the Feynman rules contain also a non-trivial braiding rule for 
each crossing, which comes from a flatness condition in a graph of intersecting particles \cite{Freidel:2005bb}. This can be incorporated as a braiding between the scalar fields, 
\begin{equation}
\psi(\phitilde{g_1}\phitilde{g_2})=\phitilde{g_2}\phitilde{g_2^{-1}g_1g_2}, \label{braid3dnoncom}
\end{equation}
in the braided quantum field theory.

From the direct analysis of the Feynman graphs with this braiding rule, one can easily find
that the energy-momentum mentioned above is conserved also in the non-planar diagrams. 
This suggests the existence of a translational symmetry in the quantum field theory.
In the sequel, we will discuss the embedding of this field theory into the framework
of braided quantum field theory,
and will check the four conditions for its translational and rotational symmetries.
 
We use the momentum representation, and take $X$ as the space of $\phitilde{g}$ 
and $\X$ as that of $\frac{\delta}{\delta \phitilde{g}}$. 
We take the braided Hopf algebra of the fields as follows,
\begin{align}
\Delta &: \phitilde{g}\to \phitilde{g}\otimesh \1+\1\otimesh \phitilde{g}, \\
\epsilon &: \phitilde{g} \to 0, \\
S &: \phitilde{g}\to -\phitilde{g}, \\
\psi &: \phitilde{g_1}\otimes \phitilde{g_2}\to \phitilde{g_2}\otimes \phitilde{g_2^{-1}g_1g_2}.
\end{align}

The evaluation and coevaluation maps are given by
\begin{align}
{\rm ev}&: \frac{\delta}{\delta \tilde \phi(g)} \otimes \tilde \phi(g')\rightarrow \delta(g^{-1}g'), \\
{\rm coev}&: 1\rightarrow \int dg \tilde \phi(g) \otimes \frac{\delta}{\delta \tilde \phi(g)}.
\label{eq:evcoev}
\end{align}
 
From $\gamma (\partial )=\partial S_0=(P^2(g)-m^2)\phitilde{g^{-1}}$,
\begin{equation}
\gamma^{-1}(\phitilde{g})=\frac{1}{P^2(g^{-1})-m^2}\frac{\delta}{\delta \phitilde{g^{-1}}}.
\end{equation}

\begin{figure}
\begin{center}
\includegraphics[scale=0.7]{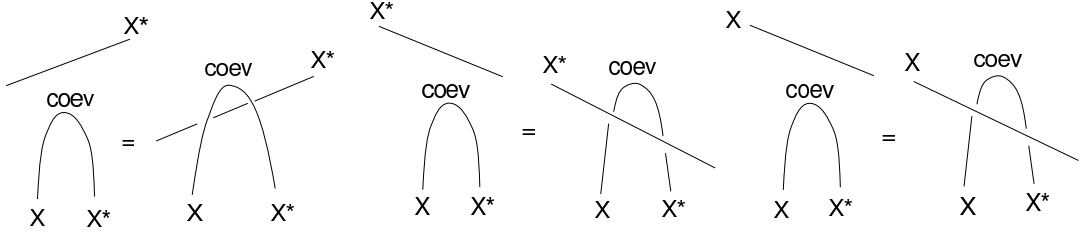}
\end{center}
\caption{The algebraic consistency conditions of coevaluation map and $X$, $\X$.}
\label{fig:consistencycond}
\end{figure}

From the algebraic consistencies in Figure \ref{fig:consistencycond}, the braidings between $X$ and $\X$ and the braiding between $\X$s are determined to be 
\begin{align}
\psi \bigg(\frac{\delta}{\delta \phitilde{g_1}}\otimes \phitilde{g_2} \bigg)&=\phitilde{g_2}\otimes \frac{\delta}{\delta \phitilde{g_2^{-1}g_1g_2}}, \\
\psi \bigg(\phitilde{g_1}\otimes \frac{\delta}{\delta \phitilde{g_2}} \bigg)&=\frac{\delta}{\delta \phitilde{g_2}}\otimes \phitilde{g_2g_1g_2^{-1}} ,\\
\psi \bigg(\frac{\delta}{\delta \phitilde{g_1}}\otimes \frac{\delta}{\delta \phitilde{g_2}} \bigg)&=\frac{\delta}{\delta \phitilde{g_2}}\otimes \frac{\delta}{\delta \phitilde{g_2g_1g_2^{-1}}}.
\end{align}
In this derivation, we have used the invariance of the Haar measure
$d(g^{-1}g'g)=dg'$.

Now we consider a translational transformation of the field. If we shift $x^i$ to $x^i+\epsilon^i$, a field $\phi(x)$ becomes
\begin{align}
\phi(x)&\to \phi(x+\epsilon) \notag \\
&=\int dg \phitilde{g}e^{i(x+\epsilon)^iP_i(g)}\notag \\
&\sim \int dg (1+i\epsilon^iP_i(g))\phitilde{g}e^{ix^iP_i(g)}.
\end{align}
Thus in the momentum representation, the translational transformation corresponds to an \textit{action} 
\begin{equation}
P^i\tar \phitilde{g} = P^i(g)\phitilde{g},~~
P^0\tar \phitilde{g} = P^0(g)\phitilde{g}. \label{eq:translop}
\end{equation}
From the requirement that the star product (\ref{eq:defofstar}) conserve a kind of momentum, 
the \textit{action} on a product of fields should be
\begin{align}
P^i\tar (\phitilde{g_1}\phitilde{g_2})&= P^i(g_1g_2)\phitilde{g_1}\phitilde{g_2} \notag \\
&=(P_1^0P_2^i+P^0_2P_1^i+\kappa \epsilon^{ijk}P_1^jP_2^k)\phitilde{g_1}\phitilde{g_2}, \label{eq:piaction} \\
P^0\tar (\phitilde{g_1}\phitilde{g_2})&=(P_1^0P_2^0-\kappa^2P_1^iP_{2i})\phitilde{g_1}\phitilde{g_2}. \label{eq:p0action}
\end{align}
This determines the coproduct of $P^i, P^0$ as 
\begin{align}
\Delta' (P^i)&=P^0 \otimes P^i+P^i\otimes P^0+\kappa \epsilon^{ijk}P^j\otimes P^k, \\
\Delta' (P^0)&=P^0\otimes P^0-\kappa^2P^i\otimes P_i.
\end{align}
This coproduct satisfies the coassociativity, which essentially comes from the 
associativity of the group multiplication.

From the axiom (\ref{eq:axiomcounit}), the counit of $P^i, P^0$ is given by
\begin{equation}
\epsilon'(P^i)=\epsilon'(P^0)=0.
\end{equation}

Since the conservation of momentum under the coevaluation map (\ref{eq:evcoev})
requires that the \textit{action} of $P^i$ on $\int dg (\phitilde{g}\otimes \frac{\delta}{\delta \phitilde{g}})$ vanish from (\ref{eq:unitaction}), 
the \textit{action} of $P^i$ on $\frac{\delta}{\delta \phitilde{g}}$ must be
\begin{equation}
P^i\tar \frac{\delta}{\delta \phitilde{g}}= P^i(g^{-1})\frac{\delta}{\delta \phitilde{g}}.
\end{equation}

In the following, we 
see that the momentum algebra satisfies the four conditions (\ref{eq:cond1}), (\ref{eq:cond2}), (\ref{eq:cond3}), (\ref{eq:cond4}).

Condition 1 is satisfied since
\begin{align}
&P^i\tar S_{int} \notag \\
= &\int dg_1dg_2dg_3\delta(g_1g_2g_3)P^i\tar (\phitilde{g_1}\phitilde{g_2}\phitilde{g_3}) \notag \\
=&\int dg_1dg_2dg_3\delta(g_1g_2g_3)P^i(g_1g_2g_3)(\phitilde{g_1}\phitilde{g_2}\phitilde{g_3}) \notag \\
=&~0.
\end{align}

Condition 2 is satisfied since
\begin{align}
\psi ( P^i\tar (\phitilde{g_1} \phitilde{g_2}))&=P^i(g_1g_2)(\phitilde{g_2} \phitilde{g_2^{-1}g_1g_2}), \notag \\
P^i\tar \psi (\phitilde{g_1} \phitilde{g_2})&=P^i(g_2g_2^{-1}g_1g_2)(\phitilde{g_2} \phitilde{g_2^{-1}g_1g_2}). \notag
\end{align}

Condition 3 is satisfied since
\begin{align}
P^i\tar \gamma^{-1}(\phitilde{g})&=\frac{1}{P^2(g^{-1})-m^2}P^i(g)\frac{\delta}{\delta \phitilde{g^{-1}}}, \notag \\
\gamma^{-1}(P^i \tar \phitilde{g})&=\frac{1}{P^2(g^{-1})-m^2}P^i(g)\frac{\delta}{\delta \phitilde{g^{-1}}}. \notag
\end{align}

Condition 4 is satisfied since
\begin{align}
\ev \bigg(P^i\tar \bigg(\frac{\delta}{\delta \phitilde{g_1}}\otimes \phitilde{g_2}\bigg)\bigg)&=P^i(g_1^{-1}g_2)~\ev \bigg(\frac{\delta}{\delta \phitilde{g_1}}\otimes \phitilde{g_2}\bigg) \notag \\
&=0.
\end{align}

Thus we find that the effective braided noncommutative field theory of three-dimensional quantum gravity coupled with spinless particles has the translational symmetry.

Next we consider a rotational symmetry. The rotational symmetry corresponds to an \textit{action}
\begin{equation}
\Lambda \tar \phitilde{g}= \phitilde{h^{-1}gh},
\end{equation}
which is the usual Lie-group one.
The \textit{action} on the tensor product is 
\begin{equation}
\Lambda \tar (\phitilde{g_1}\otimes \phitilde{g_2})= \phitilde{h^{-1}g_1h}\otimes \phitilde{h^{-1}g_2h}.
\end{equation}
Thus the coproduct of the rotational symmetry is given by
\begin{equation}
\Delta' (\Lambda) =\Lambda\otimes \Lambda.
\end{equation}
From the axiom (\ref{eq:axiomcounit}), the counit of $\Lambda$ is given by
\begin{equation}
\epsilon'(\Lambda)=1.
\end{equation}

Condition 1 is satisfied since
\begin{align}
&\Lambda \tar S_{int} \notag \\
=&\int dg_1dg_2dg_3\delta(g_1g_2g_3)\Lambda \tar (\phitilde{g_1}\phitilde{g_2}\phitilde{g_3}) \notag \\
=&\int dg_1dg_2dg_3\delta(g_1g_2g_3)(\phitilde{h^{-1}g_1h}\phitilde{h^{-1}g_2h}\phitilde{h^{-1}g_3h}) \notag \\
=&\epsilon'(\Lambda)S_{int}.
\end{align}

Condition 2 is satisfied since
\begin{align}
\psi (\Lambda \tar (\phitilde{g_1}\otimes \phitilde{g_2}))&=\phitilde{h^{-1}g_2h}\otimes \phitilde{h^{-1}g_2^{-1}g_1g_2h} \notag \\
\Lambda \tar \psi (\phitilde{g_1}\otimes \phitilde{g_2})&=\phitilde{h^{-1}g_2h}\otimes \phitilde{h^{-1}g_2^{-1}g_1g_2h}.
\end{align}

Condition 3 is satisfied since
\begin{align}
\Lambda \tar \gamma^{-1}(\phitilde{g})&=\frac{1}{P^2(g^{-1})-m^2}\frac{\delta}{\delta \phitilde{h^{-1}g^{-1}h}} \notag \\
\gamma^{-1}(\Lambda \tar \phitilde{g})&=\frac{1}{P^2(h^{-1}g^{-1}h)-m^2}\frac{\delta}{\delta \phitilde{h^{-1}g^{-1}h}} \notag \\
&=\frac{1}{P^2(g^{-1})-m^2}\frac{\delta}{\delta \phitilde{h^{-1}g^{-1}h}}.
\end{align}

Condition 4 is satisfied since
\begin{align}
\ev \bigg(\Lambda \tar \bigg(\frac{\delta}{\delta \phitilde{g_1}}\otimes \phitilde{g_2}\bigg)\bigg)&=\ev \bigg(\frac{\delta}{\delta \phitilde{h^{-1}g_1h}}\otimes \phitilde{h^{-1}g_2h}\bigg) \notag \\
&=\delta(g_1^{-1}g_2) \notag \\
&=\epsilon'(\Lambda)\ev\bigg(\frac{\delta}{\delta \phitilde{g_1}}\otimes \phitilde{g_2}\bigg).
\end{align}

Thus we find that this braided noncommutative field theory has also the rotational symmetry.


\subsection{Twisted Poincar\'e symmetry of noncommutative field theory on Moyal plane} \label{subsec:twistednc}
In this subsection, we discuss the twisted Poincar\'e symmetry of noncommutative field theory on Moyal plane $[x^{\mu},x^{\nu}]=i\theta^{\mu\nu}$. 

For example, the action of a $\phi^3$ theory is given by
\begin{equation}
S=\int d^Dx \bigg[\frac{1}{2}(\partial_{\mu}\phi \ast \partial^{\mu}\phi)(x)-\frac{1}{2}m^2(\phi \ast \phi)(x)+\frac{\lambda}{3!}(\phi \ast \phi \ast \phi)(x) \bigg],
\end{equation}
where the star product is given by
\begin{equation}
\phi(x)\ast \phi(x)=e^{\frac{i}{2}\theta^{\mu\nu}\partial^x_{\mu}\partial^y_{\nu}}\phi(x)\phi(y)\Big|_{x=y}.
\end{equation}
In the momentum representation, the action is
\begin{align}
S&=\int d^Dp \bigg[\frac{1}{2}(p^2-m^2)\phitilde{p}\phitilde{-p} \notag \\
&+\frac{\lambda}{3!}\int d^Dp_1d^Dp_2d^Dp_3e^{-\frac{i}{2}p_{1\mu}\theta^{\mu\nu}p_{2\nu}}\delta(p_1+p_2+p_3)\phitilde{p_1}\phitilde{p_2}\phitilde{p_3}\bigg].
\end{align}
We take $X$ as the space of $\phitilde{p}$ and $\X$ as that of $\frac{\delta}{\delta \phitilde{p}}$. Then we take the braided Hopf algebra as follows:
\begin{align}
\Delta &: \phitilde{p}\to \phitilde{p}\otimesh \1+\1\otimesh \phitilde{p} ,\\
\epsilon &: \phitilde{p} \to 0 ,\\
S &: \phitilde{p} \to -\phitilde{p}.
\end{align}
From $\gamma (\partial )=\partial S_0=(p^2-m^2)\phitilde{-p}$,
\begin{equation}
\gamma^{-1}(\phitilde{p})=\frac{1}{p^2-m^2}\frac{\delta}{\delta \phitilde{-p}}.
\end{equation}

Let us consider the twisted Poincar\'{e} symmetry 
\cite{Chaichian:2004za,Wess:2003da,Koch:2004ud}. 
The coproduct and the counit of the twisted Poincar\'{e} algebra is given by
\begin{align}
\Delta' (P^{\mu})&=P^{\mu}\otimes \1+\1\otimes P^{\mu}, \notag \\
\epsilon' (P^{\mu})&=0, \notag \\
\Delta' (M^{\mu\nu})&=M^{\mu\nu}\otimes \1+\1\otimes M^{\mu\nu} \notag \\
&-\frac{1}{2}\theta^{\alpha\beta}[(\delta_{\alpha}^{\mu}P^{\nu}-\delta_{\alpha}^{\nu}P^{\mu})\otimes P_{\beta}+P_{\alpha}\otimes (\delta_{\beta}^{\mu}P^{\nu}-\delta_{\beta}^{\nu}P^{\mu})], \notag \\
\epsilon'(M^{\mu\nu})&=0.
\end{align}
Thus the \textit{action} of the twisted Lorentz algebra on the tensor product is 
\begin{align}
M^{\mu\nu}\tar (\phitilde{p_1}\otimes \phitilde{p_2})&= M^{\mu\nu}\tar \phitilde{p_1}\otimes \phitilde{p_2}+\phitilde{p_1}\otimes M^{\mu\nu}\tar \phitilde{p_2} \notag \\
&-\frac{1}{2}\theta^{\alpha \beta}[(\delta_{\alpha}^{\mu}P^{\nu}-\delta_{\alpha}^{\nu}P^{\mu})\tar \phitilde{p_1}\otimes P_{\beta}\tar \phitilde{p_2} \notag \\
&+P_{\alpha}\tar \phitilde{p_1}\otimes (\delta_{\beta}^{\mu}P^{\nu}-\delta_{\beta}^{\nu}P^{\mu})\tar \phitilde{p_2}], \label{eq:transcoprotheta}
\end{align}
where $M^{\mu\nu}\tar \phitilde{p}=i(p^{\mu}\partial /\partial p_{\nu}-p^{\nu}\partial /\partial p_{\mu})\phitilde{p}$ and $P^{\mu}\tar \phitilde{p}=p^{\mu}\phitilde{p}$. The \textit{actions} of $M^{\mu\nu}$ and $P^{\mu}$ on $\frac{\delta}{\delta \phitilde{p}}$ are
\begin{align}
M^{\mu\nu}\tar \frac{\delta}{\delta \phitilde{p}}&=i(p^{\mu}\partial /\partial p_{\nu}-p^{\nu}\partial /\partial p_{\mu})\frac{\delta}{\delta \phitilde{p}}, \\
P^{\mu}\tar \frac{\delta}{\delta \phitilde{p}}&=-p^{\mu}\frac{\delta}{\delta \phitilde{p}}.
\end{align}

One easily finds that three conditions (\ref{eq:cond1}), (\ref{eq:cond3}), (\ref{eq:cond4}) are satisfied, but (\ref{eq:cond2}) is not if the braiding is trivial. 
In order to keep the invariance, the braiding must be taken as 
\begin{equation}
\psi(\phitilde{p_1}\otimes \phitilde{p_2})=e^{i\theta^{\alpha \beta}p_{2\alpha}\otimes p_{1\beta}}(\phitilde{p_2}\otimes \phitilde{p_1}).
\end{equation}
This agrees with the previous proposal \cite{Balachandran:2005eb,Oeckl:2000eg}.

We can easily check that the translational symmetry holds since the coproduct $\Delta' (P^{\mu})$ follows the usual Leibniz rule.

\subsection{Relations among correlation functions : Examples}
Now we have checked, in all orders of perturbation, that the two theories in the preceding sections have symmetry relations among correlation functions implied by 
the Hopf algebra symmetries.
In Section \ref{subsec:3dgravity} we gave how the translational generator acts on a product of fields in (\ref{eq:piaction}), (\ref{eq:p0action}) in the momentum representation. 
Since the physical meaning of this Hopf algebra transformation is not so clear, 
it would be interesting to see explicitly the symmetry relations
among correlation functions.
The same thing is also true in the case of the twisted Lorentz symmetry in Section \ref{subsec:twistednc}. In this subsection, we work out explicitly some relations among correlation functions in the two theories.

In the effective quantum field theory of quantum gravity,
the \textit{action} of the translational generators on a correlation function is given by
\begin{equation}
\langle \phitilde{g_1}\cdots \phitilde{g_n} \rangle\to i\epsilon^iP_i(g_1\cdots g_n)\langle \phitilde{g_1}\cdots \phitilde{g_n} \rangle
\end{equation}
in the momentum representation, where $\epsilon^i$ is an infinitesimal parameter.  Thus we
obtain a relation,
\begin{equation}
\label{eq:wtmom}
P_i(g_1\cdots g_n)\langle \phitilde{g_1}\cdots \phitilde{g_n} \rangle=0.
\end{equation}
This is a (modified) momentum conservation law; the correlation function has support
only on the vanishing momentum subspace, $P_i(g_1\cdots g_n)=0$.
This all-order relation in the quantum field theory would be a simple but an important implication of the Hopf algebraic translational symmetry. 
This provides a good example of the physical importance of a Hopf algebraic symmetry:
a Hopf algebra symmetry leads to a (modified) conservation law.

It would also be interesting to see the relations in the coordinate representations,
where the fields are defined by $\phi(x)=\int_p e^{ip\cdot x}\phitilde{p}$. 
As explicitly noted in the preceding subsections, 
we stress that the basis of the spaces $X$ of the field variables 
in the path integrals are parameterized in terms of momenta, and that $\phi(x)$ are 
defined by some $c$-number linear combinations of them. Therefore, an {\it action}
$a\in \mathcal{A}$ of a symmetry transformation acts as 
\begin{equation}
a\tar \phi(x)=\int_p e^{ip\cdot x} (a\tar \phitilde{p}),
\end{equation}
and the symmetry relations of correlation functions can be obtained by some inverse
Fourier transformations (with possible non-trivial measures) 
of those in momentum representations.

For example, in the case of the two point function, after the inverse Fourier transformation, the relation among correlation functions is given by 
\begin{equation}
\langle \partial^i\phi(x_1)\phi(x_2)+\phi(x_1)\partial^i\phi(x_2)\rangle=0,
\end{equation}
where we have used the relation (\ref{eq:wtmom}). Interestingly, 
this is the usual relation in a translationally invariant quantum field theory.  
In the case of the three point function, however, the relation is given by
\begin{align}
&\langle \partial^i\phi(x_1)\sqrt{1+\kappa^2 \partial^2}\phi(x_2)\sqrt{1+\kappa^2 \partial^2}\phi(x_3)+\sqrt{1+\kappa^2 \partial^2}\phi(x_1)\partial^i\phi(x_2)\sqrt{1+\kappa^2 \partial^2}\phi(x_3) \notag \\
& +\sqrt{1+\kappa^2 \partial^2}\phi(x_1)\sqrt{1+\kappa^2 \partial^2}\phi(x_2)\partial^i\phi(x_3)+i\kappa\epsilon^{ijk}\sqrt{1+\kappa^2 \partial^2}\phi(x_1)\partial_j\phi(x_2)\partial_k\phi(x_3)\notag \\
&+i\kappa\epsilon^{ijk}\partial_j\phi(x_1)\sqrt{1+\kappa^2 \partial^2}\phi(x_2)\partial_k\phi(x_3) -i\kappa\epsilon^{ijk}\partial_j\phi(x_1)\partial_k\phi(x_2)\sqrt{1+\kappa^2 \partial^2}\phi(x_3)\notag \\
&+\kappa^2\partial_j\phi(x_1)\partial^j\phi(x_2)\partial^i\phi(x_3)-\kappa^2\partial^i\phi(x_1)\partial_k\phi(x_2)\partial^k\phi(x_3) \notag \\
&+\kappa^2\partial_k\phi(x_1)\partial^i\phi(x_2)\partial^k\phi(x_3)\rangle=0.
\end{align}
This is quite a non-trivial relation among correlation functions, and would be hard to
find, if the Hopf algebra symmetry in the quantum field theory was not noticed. This 
would be another interesting 
example implying the physical importance of a Hopf algebra symmetry.  
In general, the relation has the form, 
\begin{align}
(\sum_{l=1}^n\partial_{x_li}-i\sum_{l<m}^n\kappa \epsilon^{ijk}\partial_{x_lj}\partial_{x_mk}+\mathcal{O}(\kappa^2))\langle \phi(x_1)\cdots \phi(x_n)\rangle=0.
\end{align}
In the $\kappa \to 0$ limit, the relation approaches the usual relation.
Thus the Hopf algebra symmetry is a kind of translational symmetry 
modified by adding $\kappa$ dependent higher derivative 
multi-field contributions.

We can proceed in a similar manner for the twisted Lorentz symmetry. We have a general form of such a symmetry relation as 
\begin{equation}
M_{\mu\nu} \tar \langle \phitilde{p_1}\cdots \phitilde{p_n}\rangle=0.
\label{eq:generalmmunu}
\end{equation}
In the case of the two point function, the relation is given by
\begin{align}
\langle (x_1^{\mu}\partial^{\nu}-x_1^{\nu}\partial^{\mu})\phi(x_1)\phi(x_2)+\phi(x_1)(x_2^{\mu}\partial^{\nu}-x_2^{\nu}\partial^{\mu})\phi(x_2)\rangle=0, \label{eq:2pointxrep}
\end{align}
where we have used the momentum conservation. This is the same relation as that in 
a Lorentz invariant quantum field theory. 
In the case of the three point function, the relation is given by
\begin{align}
&\langle (x_1^{\mu}\partial^{\nu}-x_1^{\nu}\partial^{\mu})\phi(x_1)\phi(x_2)\phi(x_3) \notag \\
&+\phi(x_1)(x_2^{\mu}\partial^{\nu}-x_2^{\nu}\partial^{\mu})\phi(x_2)\phi(x_3)
+\phi(x_1)\phi(x_2)(x_3^{\mu}\partial^{\nu}-x_3^{\nu}\partial^{\mu})\phi(x_3) \notag \\
&+\frac{1}{2} i\theta^{\alpha \mu}(\partial_{\alpha}\phi(x_1)\partial^{\nu}\phi(x_2)\phi(x_3)+\partial_{\alpha}\phi(x_1)\phi(x_2)\partial^{\nu}\phi(x_3)+\phi(x_1)\partial_{\alpha}\phi(x_2)\partial^{\nu}\phi(x_3) \notag \\
&-\partial^{\nu}\phi(x_1)\partial_{\alpha}\phi(x_2)\phi(x_3)-\partial^{\nu}\phi(x_1)\phi(x_2)\partial_{\alpha}\phi(x_3)-\phi(x_1)\partial^{\nu}\phi(x_2)\partial_{\alpha}\phi(x_3)) \notag \\
&-\frac{1}{2} i\theta^{\alpha \nu}(\partial_{\alpha}\phi(x_1)\partial^{\mu}\phi(x_2)\phi(x_3)+\partial_{\alpha}\phi(x_1)\phi(x_2)\partial^{\mu}\phi(x_3)+\phi(x_1)\partial_{\alpha}\phi(x_2)\partial^{\mu}\phi(x_3) \notag \\
&-\partial^{\mu}\phi(x_1)\partial_{\alpha}\phi(x_2)\phi(x_3)-\partial^{\mu}\phi(x_1)\phi(x_2)\partial_{\alpha}\phi(x_3)-\phi(x_1)\partial^{\mu}\phi(x_2)\partial_{\alpha}\phi(x_3))\rangle=0. \label{eq:3pointxrep}
\end{align}
In general, the relation among correlation functions has the from, 
\begin{equation}
((x_{1\mu}\partial_{x_1\nu}-x_{1\nu}\partial_{x_1\mu})+\cdots +(x_{n\mu}\partial_{x_n\nu}-x_{n\nu}\partial_{x_m\nu})+\mathcal{O}(\theta))\langle \phi(x_1)\cdots \phi(x_n)\rangle=0
\end{equation}
in the coordinate representation. The leading terms corresponds to the usual Lorentz transformation $x^{\mu}\to x^{\mu}+\epsilon^{\mu\nu}x_{\nu}$. 

The above symmetry relations on Moyal plane can be represented in similar manners as the usual commutative cases, if we use star products. 
In the papers \cite{Tureanu:2006pb, Zahn:2006wt, Bu:2006ha, Abe:2006ig, Balachandran:2006pi, Fiore:2007vg, Joung:2007qv}, they have pointed out that in coordinate representation, correlation functions on Moyal plane should be defined with star products extended to non-coincident points (see also \cite{Szabo:2001kg}) instead of usual products since the usual commutative commutation relation $[x_i^{\mu}, x_j^{\nu}]=0 ~(i,j=1, \cdots, n)$ is not invariant under the twisted Poincar\'{e} transformation. Carrying out Fourier transformation of the symmetry relation 
(\ref{eq:generalmmunu}) in momentum representation to such a noncommutative coordinate representation, we obtain the symmetry relations in star tensor products.  
Namely (\ref{eq:2pointxrep}) becomes
\begin{align}
\langle ((x_1^{\mu}\partial^{\nu}-x_1^{\nu}\partial^{\mu})\phi(x_1))\ast \phi(x_2)+\phi(x_1)\ast ((x_2^{\mu}\partial^{\nu}-x_2^{\nu}\partial^{\mu})\phi(x_2))\rangle=0,
\end{align}
and (\ref{eq:3pointxrep}) becomes
\begin{align}
&\langle ((x_1^{\mu}\partial^{\nu}-x_1^{\nu}\partial^{\mu})\phi(x_1))\ast \phi(x_2)\ast \phi(x_3) \notag \\
&+\phi(x_1)\ast ((x_2^{\mu}\partial^{\nu}-x_2^{\nu}\partial^{\mu})\phi(x_2))\ast \phi(x_3)
+\phi(x_1)\ast \phi(x_2)\ast ((x_3^{\mu}\partial^{\nu}-x_3^{\nu}\partial^{\mu})\phi(x_3)) \notag \\
&+\frac{1}{2} i\theta^{\alpha \mu}(\partial_{\alpha}\phi(x_1)\ast \partial^{\nu}\phi(x_2)\ast \phi(x_3)+\partial_{\alpha}\phi(x_1)\ast \phi(x_2)\ast \partial^{\nu}\phi(x_3)+\phi(x_1)\ast \partial_{\alpha}\phi(x_2)\ast \partial^{\nu}\phi(x_3) \notag \\
&-\partial^{\nu}\phi(x_1)\ast \partial_{\alpha}\phi(x_2)\ast \phi(x_3)-\partial^{\nu}\phi(x_1)\ast \phi(x_2)\partial_{\alpha}\ast \phi(x_3)-\phi(x_1)\ast \partial^{\nu}\phi(x_2)\ast \partial_{\alpha}\phi(x_3)) \notag \\
&-\frac{1}{2} i\theta^{\alpha \nu}(\partial_{\alpha}\phi(x_1)\ast \partial^{\mu}\phi(x_2)\ast \phi(x_3)+\partial_{\alpha}\phi(x_1)\ast \phi(x_2)\partial^{\mu}\ast \phi(x_3)+\phi(x_1)\ast \partial_{\alpha}\phi(x_2)\ast \partial^{\mu}\phi(x_3) \notag \\
&-\partial^{\mu}\phi(x_1)\ast \partial_{\alpha}\phi(x_2)\ast \phi(x_3)-\partial^{\mu}\phi(x_1)\ast \phi(x_2)\ast \partial_{\alpha}\phi(x_3)-\phi(x_1)\ast \partial^{\mu}\phi(x_2)\ast \partial_{\alpha}\phi(x_3))\rangle=0.
\end{align}

More generally we can derive the symmetry relations of correlation functions for tensor fields $\phi_{\alpha_1\cdots \alpha_n}(x)\equiv \partial_{\alpha_1}\cdots \partial_{\alpha_n}\phi(x)$. For example in the case of the three point function of the tensor fields, the symmetry relation becomes
\begin{align}
&\langle ((M^{1\mu\nu})_{\alpha_1\cdots \alpha_l}{}^{\delta_1\cdots \delta_l}\phi_{\delta_1\cdots \delta_l}(x_1))\ast \phi_{\beta_1\cdots \beta_m}(x_2)\ast \phi_{\gamma_1\cdots \gamma_n}(x_3) \notag \\
&+\phi_{\alpha_1\cdots \alpha_l}(x_1)\ast ((M^{2\mu\nu})_{\beta_1\cdots \beta_m}{}^{\delta_1\cdots \delta_m}\phi_{\delta_1\cdots \delta_m}(x_2))\ast \phi_{\gamma_1\cdots \gamma_n}(x_3) \notag \\
&+\phi_{\alpha_1\cdots \alpha_l}(x_1)\ast \phi_{\beta_1\cdots \beta_m}(x_2)\ast ((M^{3\mu\nu})_{\gamma_1\cdots \gamma_n}{}^{\delta_1\cdots \delta_n}\phi_{\delta_1\cdots \delta_n}(x_3)) \notag \\
&-\frac{1}{2}\theta^{\alpha \mu}[\partial_{\alpha}\phi_{\alpha_1\cdots \alpha_l}(x_1)\ast \partial^{\nu}\phi_{\beta_1\cdots \beta_m}(x_2)\ast \phi_{\gamma_1\cdots \gamma_n}(x_3) \notag \\
&~~~~~~~~~~~+\partial_{\alpha}\phi_{\alpha_1\cdots \alpha_l}(x_1)\ast \phi_{\beta_1\cdots \beta_m}(x_2)\ast \partial^{\nu}\phi_{\gamma_1\cdots \gamma_n}(x_3) \notag \\
&~~~~~~~~~~~+\phi_{\alpha_1\cdots \alpha_l}(x_1)\ast \partial_{\alpha}\phi_{\beta_1\cdots \beta_m}(x_2)\ast \partial^{\nu}\phi_{\gamma_1\cdots \gamma_n}(x_3) \notag \\
&~~~~~~~~~~~-\partial^{\nu}\phi_{\alpha_1\cdots \alpha_l}(x_1)\ast \partial_{\alpha}\phi_{\beta_1\cdots \beta_m}(x_2)\ast \phi_{\gamma_1\cdots \gamma_n}(x_3) \notag \\
&~~~~~~~~~~~-\partial^{\nu}\phi_{\alpha_1\cdots \alpha_l}(x_1)\ast \phi_{\beta_1\cdots \beta_m}(x_2)\partial_{\alpha}\ast \phi_{\gamma_1\cdots \gamma_n}(x_3) \notag \\
&~~~~~~~~~~~-\phi_{\alpha_1\cdots \alpha_l}(x_1)\ast \partial^{\nu}\phi_{\beta_1\cdots \beta_m}(x_2)\ast \partial_{\alpha}\phi_{\gamma_1\cdots \gamma_n}(x_3)] \notag \\
&+\frac{1}{2}\theta^{\alpha \nu}[\partial_{\alpha}\phi_{\alpha_1\cdots \alpha_l}(x_1)\ast \partial^{\mu}\phi_{\beta_1\cdots \beta_m}(x_2)\ast \phi_{\gamma_1\cdots \gamma_n}(x_3) \notag \\
&~~~~~~~~~~~+\partial_{\alpha}\phi_{\alpha_1\cdots \alpha_l}(x_1)\ast \phi_{\beta_1\cdots \beta_m}(x_2)\partial^{\mu}\ast \phi_{\gamma_1\cdots \gamma_n}(x_3) \notag \\
&~~~~~~~~~~~+\phi_{\alpha_1\cdots \alpha_l}(x_1)\ast \partial_{\alpha}\phi_{\beta_1\cdots \beta_m}(x_2)\ast \partial^{\mu}\phi_{\gamma_1\cdots \gamma_n}(x_3) \notag \\
&~~~~~~~~~~~-\partial^{\mu}\phi_{\alpha_1\cdots \alpha_l}(x_1)\ast \partial_{\alpha}\phi_{\beta_1\cdots \beta_m}(x_2)\ast \phi_{\gamma_1\cdots \gamma_n}(x_3) \notag \\
&~~~~~~~~~~~-\partial^{\mu}\phi_{\alpha_1\cdots \alpha_l}(x_1)\ast \phi_{\beta_1\cdots \beta_m}(x_2)\ast \partial_{\alpha}\phi_{\gamma_1\cdots \gamma_n}(x_3) \notag \\
&~~~~~~~~~~~-\phi_{\alpha_1\cdots \alpha_l}(x_1)\ast \partial^{\mu}\phi_{\beta_1\cdots \beta_m}(x_2)\ast \partial_{\alpha}\phi_{\gamma_1\cdots \gamma_n}(x_3)]\rangle=0,
\end{align}
where 
\begin{align}
&(M^{\mu\nu})_{\alpha_1\cdots \alpha_n}{}^{\beta_1\cdots \beta_n}=(L^{\mu\nu})_{\alpha_1\cdots \alpha_n}{}^{\beta_1\cdots \beta_n}+(S^{\mu\nu})_{\alpha_1\cdots \alpha_n}{}^{\beta_1\cdots \beta_n} \notag \\
&(L^{\mu\nu})_{\alpha_1\cdots \alpha_n}{}^{\beta_1\cdots \beta_n}=i(x^{\mu}\partial^{\nu}-x^{\nu}\partial^{\mu})\delta_{\alpha_1}{}^{\beta_1}\cdots \delta_{\alpha_n}{}^{\beta_n} \notag \\
&(S^{\mu\nu})_{\alpha_1\cdots \alpha_n}{}^{\beta_1\cdots \beta_n}=i(\eta^{\nu\beta_1}\delta_{\{ \alpha_1}{}^{\mu}\delta_{\alpha_2}{}^{\beta_2}\cdots \delta_{\alpha_n \}}{}^{\beta_n}-\eta^{\mu\beta_1}\delta_{\{ \alpha_1}{}^{\nu}\delta_{\alpha_2}{}^{\beta_2}\cdots \delta_{\alpha_n \}}{}^{\beta_n})
\end{align}
If we bring the operators $(M^{i\mu\nu})_{\alpha_1\cdots \alpha_n}{}^{\beta_1\cdots \beta_n}~(i=1,2,3)$ out of the star products, $\theta^{\mu\nu}$ dependent terms are canceled. The final expressions are just the usual Lorentz rotations on the coordinates and the tensorial indices in the correlation functions.
This is fully consistent with the discussions in \cite{Fiore:2007vg}.

\subsection{Origin of Hopf algebra symmetries}
To study more the meaning of these additional terms, let us see 
closer the transformation properties of the star products.  
In the latter case, it is known that the $\theta^{\mu\nu}$ dependence of 
the twisted Lorentz transformation (\ref{eq:transcoprotheta}) comes from the Lorentz transformation of $\theta^{\mu\nu}$ itself \cite{Alvarez-Gaume:2006bn}.
To see this, let us consider 
an infinitesimal Lorentz transformation, $\Lambda^{\mu}{}_{\nu}=\delta^{\mu}{}_{\nu}+\epsilon^{\mu}{}_{\nu}$.
The transformation of $\theta^{\mu\nu}$ is given by  
\begin{align}
\theta^{\mu\nu}&\to \theta^{\mu\nu}+\epsilon^{\mu}{}_{\rho}\theta^{\rho\nu}+\epsilon^{\nu}{}_{\rho}\theta^{\mu\rho} \notag \\
&:=\theta^{\mu\nu}+\delta\theta^{\mu\nu}.
\label{eq:thetatrans}
\end{align}
If one considers not only the transformation of the coordinates,  $x^{'\mu}=x^{\mu}+\epsilon^{\mu\nu}x_{\nu}$,
but also (\ref{eq:thetatrans}), and assumes
that $\phi(x)\ast_{\theta}\phi(x)$ and $\phi'(x')\ast_{\theta+\delta \theta}\phi'(x')$
be equal, one obtains, after the Fourier transformation,
\begin{align}
&\tilde{\phi}'(p_1)\otimes \tilde{\phi}'(p_2) \notag \\
&=\big(1-\frac{i}{2}(\epsilon^{\mu\nu}M_{\mu\nu}\otimes \1+\1\otimes \epsilon^{\mu\nu}M_{\mu\nu}+\delta\theta^{\mu\nu}P_{\mu}\otimes P_{\nu})\big)\phitilde{p_1}\otimes \phitilde{p_2} \notag \\
&=\big(1-\frac{i}{2}\epsilon^{\mu\nu}\Delta'M_{\mu\nu}\big)\phitilde{p_1}\otimes \phitilde{p_2} ,\label{eq:identityoftwist}
\end{align}
which agrees with (\ref{eq:transcoprotheta}). This shows that
the additional part of the coproduct of $M_{\mu\nu}$ takes into 
account the transformation of the non-dynamical background parameter $\theta^{\mu\nu}$.

The former case can be discussed in a similar manner.
The definition of the star product is given by
\begin{equation}
e^{ix^iP_i(g_1)}\star_x e^{ix^iP_i(g_2)}=e^{ix^iP_i(g_1g_2)}, \label{eq:stardefx}
\end{equation}
where we have explicitly indicated the coordinate where the star product is taken.  
Then we recognize that $e^{i(x+\epsilon)^iP_i(g_1)}\star_{x+\epsilon}e^{i(x+\epsilon)^iP_i(g_2)}$ and $e^{i(x+\epsilon)^iP_i(g_1)}\star_{x}e^{i(x+\epsilon)^iP_i(g_2)}$ give distinct
values. 
Namely, if the coordinate of the star product is also shifted, 
\begin{equation}
e^{i(x+\epsilon)^iP_i(g_1)}\star_{x+\epsilon}e^{i(x+\epsilon)^iP_i(g_2)}=e^{i(x+\epsilon)^iP_i(g_1g_2)},
\label{eq:starshift}
\end{equation}
but, if not, 
\begin{equation}
e^{i(x+\epsilon)^iP_i(g_1)}\star_{x}e^{i(x+\epsilon)^iP_i(g_2)}=e^{i\epsilon^iP_i(g_1)}e^{i\epsilon^iP_i(g_2)}e^{ix^iP_i(g_1g_2)}.
\end{equation}
Therefore, 
if we take the translational transformation as (\ref{eq:starshift}), 
and carry out the same procedure in deriving (\ref{eq:noncommutativity}),
we always obtain a translational invariant commutation relation\footnote{There is a similar discussion in \cite{Agostini:2006nc}.},
\begin{equation}
[(x+\epsilon)^i,(x+\epsilon)^j]_{\star_{x+\epsilon}}=2i\kappa \epsilon^{ijk}(x+\epsilon)_k.
\end{equation}
Now, assuming that $\phi(x)\star_x \phi(x)$ and $\phi'(x')\star_{x'}\phi'(x')$ be 
equal under the translation $x^i\to x^{'i}=x^i+\epsilon^i$, we obtain, after the Fourier transformation,
\begin{equation}
\tilde{\phi}'(g_1)\tilde{\phi}'(g_2)=(1-i\epsilon^iP_i(g_1g_2))\phitilde{g_1}\phitilde{g_2},
\end{equation}
which is the same as (\ref{eq:piaction}). 

From these two examples, we anticipate that the multi-field contributions in (\ref{eq:wtidgeneral}) comes from the transformation properties of the star products.

\section{Summary and comments}
We have discussed symmetries in noncommutative field theories in the framework of 
braided quantum field theory. 
We have obtained the algebraic conditions for a Hopf algebra to be a symmetry
of a braided quantum field theory, by discussing the conditions for the relations among correlation functions generated from the transformation algebra to hold. 
Then we have applied our discussions to the Poincar\'{e} symmetries in the effective 
noncommutative
field theory of three-dimensional quantum gravity coupled with spinless particles 
and in the noncommutative field theory on Moyal plane. 
In the former case we can understand the braiding between fields, 
which was derived from the three-dimensional quantum gravity computation, 
from the viewpoint of the translational symmetry of the noncommutative field theory on a Lie-algebraic noncommutative spacetime. 
In the latter case we have found that the twisted Lorentz symmetry on Moyal plane is a symmetry of the quantum field theory only after the inclusion of the nontrivial braiding factor, 
which is in agreement with the previous proposal \cite{Balachandran:2006pi,Oeckl:2000eg}. 
Then we have discussed the meaning of the 
Hopf algebra symmetries from the viewpoint of coordinate representation.

In the recent research a noncommutative field theory on $\kappa$-Minkowski spacetime is discussed \cite{Freidel:2006gc}. Since this noncommutativity of the coordinates is given by $[x^0,x^j]=\frac{i}{\kappa}x^j$, this noncommutative field theory will not have the naive
translational symmetry. We may introduce a non-trivial braiding between fields as 
in the effective field theory 
discussed in Section \ref{subsec:3dgravity} to keep the momentum conservation. 
However, while the effective field theory has the braided category structure because of the 
invariance of the Haar measure $d(g^{-1}g'g)=dg'$, 
the measure of the momentum space of the field theory on $\kappa$-Minkowski spacetime is only left-invariant \cite{Freidel:2006gc}. Therefore it is not clear to us whether we can embed this field theory 
on $\kappa$-Minkowski spacetime into the framework of braided quantum field theory.

\section*{Acknowledgments}

We would like to thank S.~Terashima and S.~Sasaki for useful discussions and comments,
and would also like to thank L.~Freidel for stimulating discussions and 
explaining their recent results during his stay in Yukawa Institute for Theoretical Physics after the 21st Nishinomiya-Yukawa Memorial Symposium. 
Y.S. was supported in part by JSPS Research Fellowships for Young Scientists.
N.S. was supported in part by the Grant-in-Aid for Scientific Research No.13135213, No.16540244 and No.18340061
from the Ministry of Education, Science, Sports and Culture of Japan.

\newpage
\appendix
\section{The proofs of the formula (\ref{eq:BLR1}), (\ref{eq:BLR2})} \label{sec:app1}
We give the proofs of the formula (\ref{eq:BLR1}), (\ref{eq:BLR2}) using diagrams. At first we use the formula
\begin{equation}
\widehat{\ev}(\partial \otimes \alpha \beta)=\widehat{\ev}(\partial \otimes \alpha )\epsilon(\beta )+\widehat{\ev}(\partial \otimes \beta )\epsilon(\alpha ), \label{eq:lemma}
\end{equation}
where $\alpha, \beta \in \Xh$. This is clear from the definition of 
$\widehat{\ev}$.

\begin{figure}
\begin{center}
\includegraphics[scale=0.8]{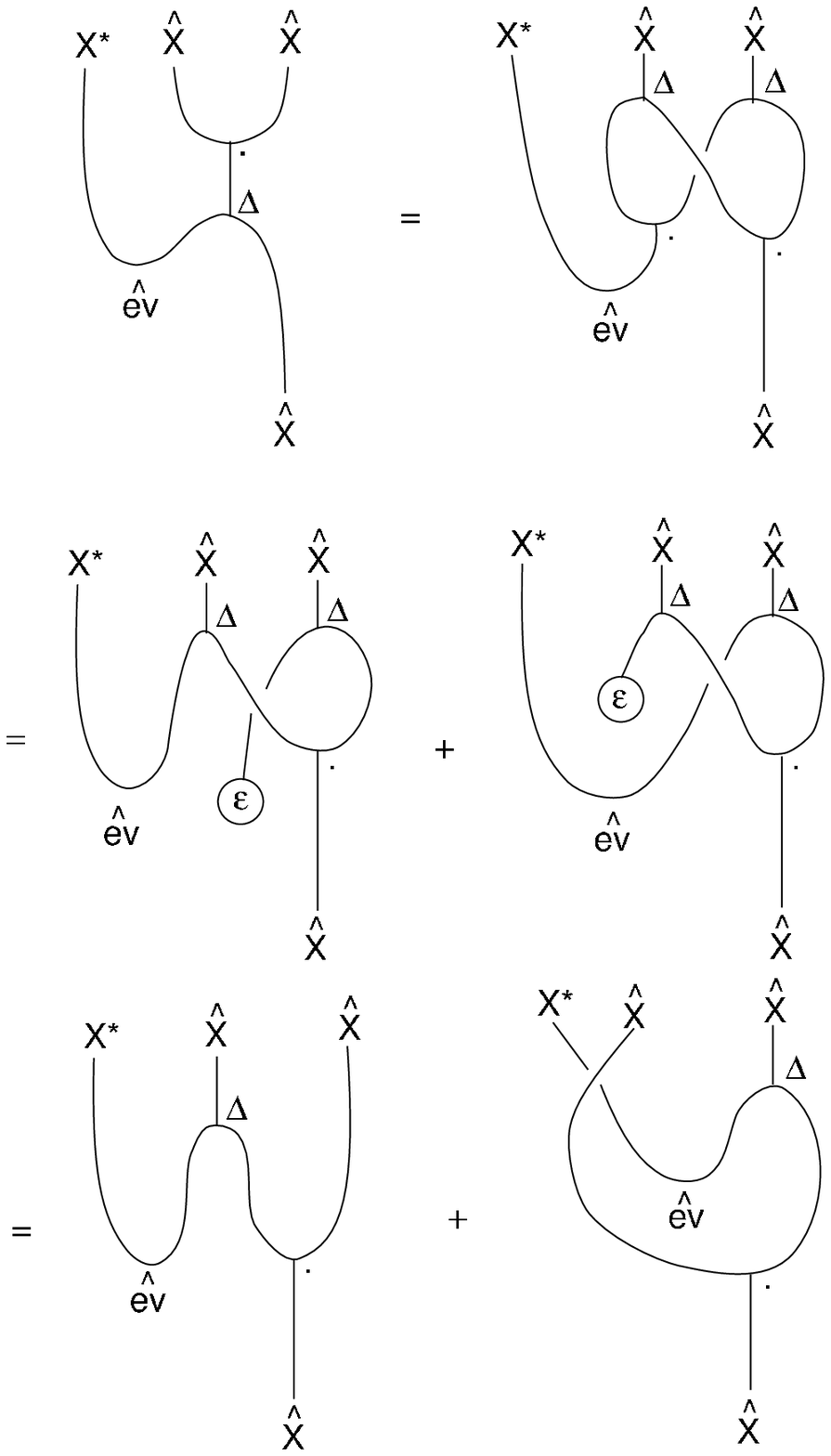}
\end{center}
\caption{The proof of (\ref{eq:BLR1}).}
\label{fig:proofofBLR}
\end{figure}

Figure \ref{fig:proofofBLR} gives the proof of (\ref{eq:BLR1}). In the first line, we use the axiom (\ref{eq:coproaxm}), and in the second line we use the lemma (\ref{eq:lemma}). We find the last line from the property of counit.

\begin{figure}
\begin{center}
\includegraphics[scale=0.8]{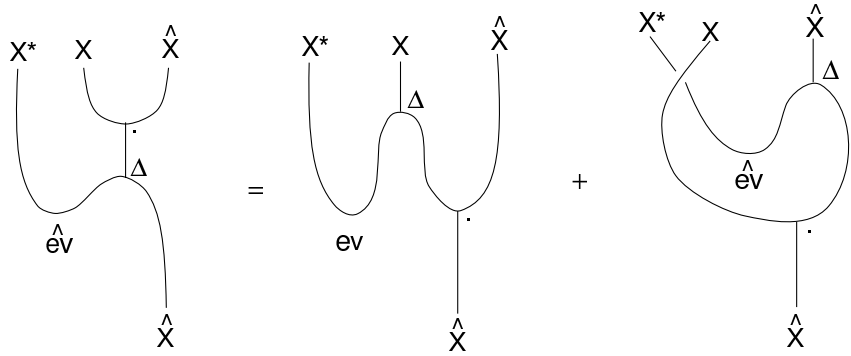}
\end{center}
\caption{The left-hand side of (\ref{eq:BLR2}).}
\label{fig:blr21}
\end{figure}

Next we prove (\ref{eq:BLR2}). By using the braided Leibniz rule (\ref{eq:BLR1}) as $\alpha \in X \otimes \Xh$, the left-hand side of (\ref{eq:BLR2}) becomes Figure \ref{fig:blr21}. The first term of Figure \ref{fig:blr21} becomes $(\ev\otimes \id^{n-1})(\partial \otimes \id^n \alpha)$ by using the definition of coproduct (\ref{eq:copro}).

\begin{figure}
\begin{center}
\includegraphics[scale=0.6]{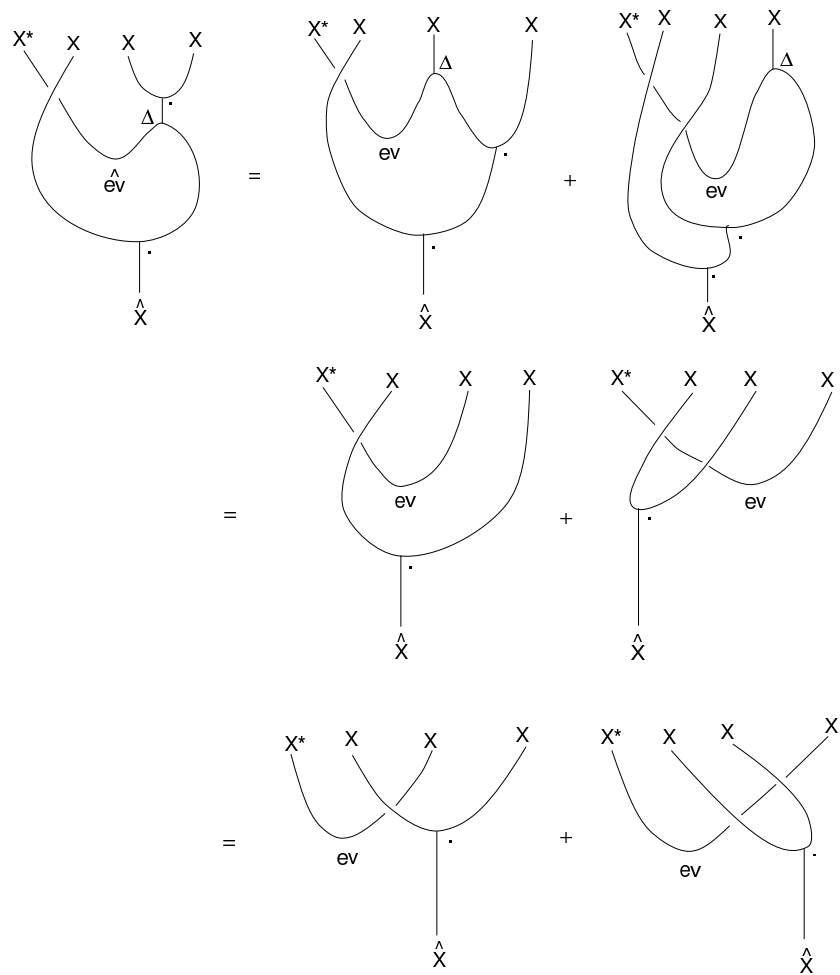}
\end{center}
\caption{The second term of Figure \ref{fig:blr21}.}
\label{fig:blr22}
\end{figure}

In the second term of Figure \ref{fig:blr21}, we divide $\Xh$ into $X \otimes \Xh$ and iterate the same as we did above. For example, if the degree of $\Xh$ is 3, the second term of Figure \ref{fig:blr21} can be reduced as in Figure \ref{fig:blr22}. We have used $\Delta X=X\otimesh \1+\1\otimesh X$ in the second line of Figure \ref{fig:blr22}. The 
result agrees with (\ref{eq:BLR2}).

In the same way, we can obtain the formula (\ref{eq:BLR2}) in general.

\section{The proofs of (\ref{eq:bwick1}), (\ref{eq:bwick2}), (\ref{eq:bwick3})} \label{sec:app2}
\begin{figure}
\begin{center}
\includegraphics[scale=0.6]{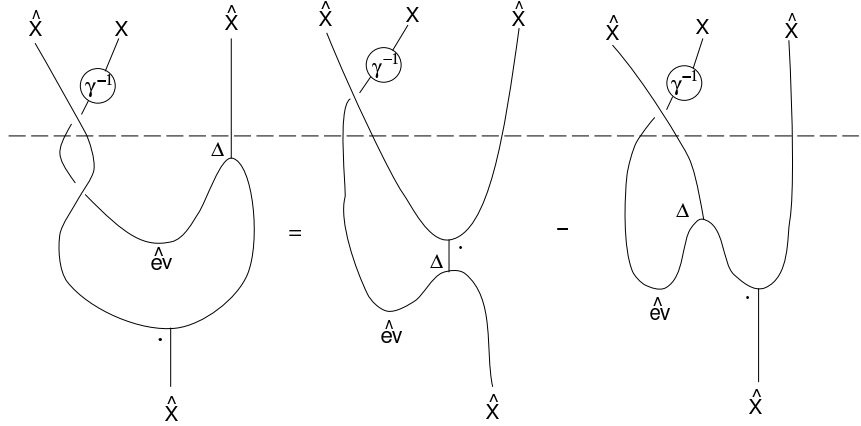}
\end{center}
\caption{The diagram obtained from adding $\gamma^{-1}$ and $\psi$ over 
the braided Leibniz rule.}
\label{fig:proofbwick1}
\end{figure}

From the definition of $\gamma $ (\ref{eq:gamma}), we find that
\begin{equation}
\alpha aw=-\alpha \diff (\gamma^{-1}(a)\otimes w), \label{eq:proofofbwick1}
\end{equation}
for $a\in X$ and $\alpha \in \Xh$. On the other hand, adding $\gamma^{-1}$ and $\psi$ to the braided Leibniz rule (\ref{eq:BLR1}) as in Figure \ref{fig:proofbwick1}, we find that
\begin{equation}
\alpha \diff (\gamma^{-1}(a)\otimes w)=\diff (\psi(\alpha\otimes \gamma^{-1}(a))w)-(\diff \circ \psi(\alpha \otimes \gamma^{-1}(a)))w. \label{eq:proofofbwick2}
\end{equation}

Combining (\ref{eq:proofofbwick1}), (\ref{eq:proofofbwick2}), we obtain 
\begin{equation}
\alpha aw=-\diff (\psi(\alpha\otimes \gamma^{-1}(a))w)+(\diff \circ \psi(\alpha \otimes \gamma^{-1}(a)))w. \label{eq:proofofbwick3}
\end{equation}

Integrating the both hand sides of (\ref{eq:proofofbwick3}) and using (\ref{eq:int}), we find that
\begin{equation}
Z^{(0)}(\alpha a)=Z^{(0)}(\diff \circ \psi (\alpha \otimes \gamma^{-1}(a))). \label{eq:eq:proofofbwick3}
\end{equation}
If $\alpha $ is $b\in X$, 
\begin{align}
Z^{(0)}(ba)&=Z^{(0)}(\diff \circ \psi (b \otimes \gamma^{-1}(a))) \notag \\
&=\ev \circ \psi(b\otimes \gamma^{-1}(a)) \notag \\
&=\ev \circ (\gamma^{-1} \otimes \id)\circ \psi(b \otimes a).
\end{align}
Thus we obtain (\ref{eq:bwick1}). 

By putting $\alpha=\1$, it is clear that 
\begin{equation}
Z^{(0)}_1(a)=0.
\end{equation}

\begin{figure}
\begin{center}
\includegraphics[scale=0.7]{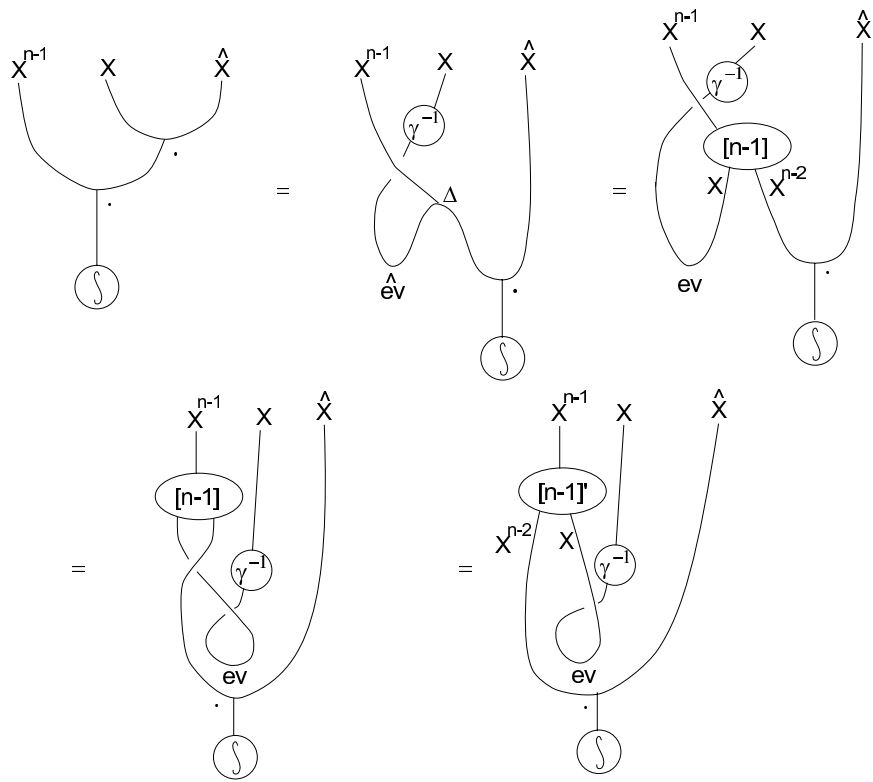}
\end{center}
\caption{Diagrammatic proof of (\ref{eq:iterationf})}
\label{fig:proofbwick2}
\end{figure}

Next we rewrite (\ref{eq:eq:proofofbwick3}) for $\alpha \in X^{n-1}$ using the formula (\ref{eq:BLR2}). Diagrammatically it is written as in Figure \ref{fig:proofbwick2}. The second equality is due to (\ref{eq:BLR2}). Thus
 we obtain that 
\begin{equation}
\Zfreed{n}=(\Zfreed{n-2}\otimes \Zfreed{2})\circ ([n-1]'_{\psi}\otimes \id) \label{eq:iterationf}
\end{equation}
Iterating this, we find (\ref{eq:bwick2}) for even $n$ and (\ref{eq:bwick3}) for odd $n$.

\newpage
\begin{thebibliography}{10}

\bibitem{Snyder:1946qz}
  H.~S.~Snyder,
  ``Quantized space-time,''
  Phys.\ Rev.\  {\bf 71}, 38 (1947).

\bibitem{Yang:1947ud}
  C.~N.~Yang,
  ``On Quantized Space-Time,''
  Phys.\ Rev.\  {\bf 72}, 874 (1947).

\bibitem{Connes:1990qp}
  A.~Connes and J.~Lott,
  ``Particle Models And Noncommutative Geometry (Expanded Version),''
  Nucl.\ Phys.\ Proc.\ Suppl.\  {\bf 18B}, 29 (1991).

\bibitem{Doplicher:1994tu}
  S.~Doplicher, K.~Fredenhagen and J.~E.~Roberts,
  ``The Quantum structure of space-time at the Planck scale and quantum fields,''
  Commun.\ Math.\ Phys.\  {\bf 172}, 187 (1995)
  [arXiv:hep-th/0303037].
    
\bibitem{Seiberg:1999vs}
  N.~Seiberg and E.~Witten,
  ``String theory and noncommutative geometry,''
  JHEP {\bf 9909}, 032 (1999)
  [arXiv:hep-th/9908142].

\bibitem{Garay:1994en}
  L.~J.~Garay,
  ``Quantum gravity and minimum length,''
  Int.\ J.\ Mod.\ Phys.\  A {\bf 10}, 145 (1995)
  [arXiv:gr-qc/9403008].


\bibitem{Sasakura:2000vc}
  N.~Sasakura,
  ``Space-time uncertainty relation and Lorentz invariance,''
  JHEP {\bf 0005}, 015 (2000)
  [arXiv:hep-th/0001161].

\bibitem{Madore:2000en}
  J.~Madore, S.~Schraml, P.~Schupp and J.~Wess,
  ``Gauge theory on noncommutative spaces,''
  Eur.\ Phys.\ J.\  C {\bf 16}, 161 (2000)
  [arXiv:hep-th/0001203].

\bibitem{Freidel:2005ec}
  L.~Freidel and S.~Majid,
  ``Noncommutative harmonic analysis, sampling theory and the Duflo map in 2+1
  quantum gravity,''
  arXiv:hep-th/0601004.


\bibitem{Imai:2000kq}
  S.~Imai and N.~Sasakura,
  ``Scalar field theories in a Lorentz-invariant three-dimensional
  noncommutative space-time,''
  JHEP {\bf 0009}, 032 (2000)
  [arXiv:hep-th/0005178].

  
\bibitem{Minwalla:1999px}
  S.~Minwalla, M.~Van Raamsdonk and N.~Seiberg,
  ``Noncommutative perturbative dynamics,''
  JHEP {\bf 0002}, 020 (2000)
  [arXiv:hep-th/9912072].

\bibitem{Chaichian:2004za}
  M.~Chaichian, P.~P.~Kulish, K.~Nishijima and A.~Tureanu,
  ``On a Lorentz-invariant interpretation of noncommutative space-time and  its
  implications on noncommutative QFT,''
  Phys.\ Lett.\  B {\bf 604}, 98 (2004)
  [arXiv:hep-th/0408069].

\bibitem{Wess:2003da}
  J.~Wess,
  ``Deformed coordinate spaces: Derivatives,''
  arXiv:hep-th/0408080.

\bibitem{Koch:2004ud}
  F.~Koch and E.~Tsouchnika,
  ``Construction of theta-Poincare algebras and their invariants on
  M(theta),''
  Nucl.\ Phys.\  B {\bf 717}, 387 (2005)
  [arXiv:hep-th/0409012].

\bibitem{Aschieri:2005yw}
  P.~Aschieri, C.~Blohmann, M.~Dimitrijevic, F.~Meyer, P.~Schupp and J.~Wess,
  ``A gravity theory on noncommutative spaces,''
  Class.\ Quant.\ Grav.\  {\bf 22}, 3511 (2005)
  [arXiv:hep-th/0504183].
  
\bibitem{Aschieri:2005zs}
  P.~Aschieri, M.~Dimitrijevic, F.~Meyer and J.~Wess,
  ``Noncommutative geometry and gravity,''
  Class.\ Quant.\ Grav.\  {\bf 23}, 1883 (2006)
  [arXiv:hep-th/0510059].  
  
\bibitem{Calmet:2005qm}
  X.~Calmet and A.~Kobakhidze,
  ``Noncommutative general relativity,''
  Phys.\ Rev.\  D {\bf 72}, 045010 (2005)
  [arXiv:hep-th/0506157].
  
\bibitem{Kobakhidze:2006kb}
  A.~Kobakhidze,
  ``Theta-twisted gravity,''
  arXiv:hep-th/0603132.  
  
\bibitem{Chaichian:2004yh}
  M.~Chaichian, P.~Presnajder and A.~Tureanu,
  ``New concept of relativistic invariance in NC space-time: Twisted  Poincare
  symmetry and its implications,''
  Phys.\ Rev.\ Lett.\  {\bf 94}, 151602 (2005)
  [arXiv:hep-th/0409096].

\bibitem{Chaichian:2005yp}
  M.~Chaichian, K.~Nishijima and A.~Tureanu,
  ``An interpretation of noncommutative field theory in terms of a quantum
  shift,''
  Phys.\ Lett.\  B {\bf 633}, 129 (2006)
  [arXiv:hep-th/0511094].
  
\bibitem{Balachandran:2005eb}
  A.~P.~Balachandran, G.~Mangano, A.~Pinzul and S.~Vaidya,
  ``Spin and statistics on the Groenwald-Moyal plane: Pauli-forbidden  levels
  and transitions,''
  Int.\ J.\ Mod.\ Phys.\  A {\bf 21}, 3111 (2006)
  [arXiv:hep-th/0508002].

\bibitem{Balachandran:2005pn}
  A.~P.~Balachandran, A.~Pinzul and B.~A.~Qureshi,
  ``UV-IR mixing in non-commutative plane,''
  Phys.\ Lett.\  B {\bf 634}, 434 (2006)
  [arXiv:hep-th/0508151].
  
\bibitem{Lizzi:2006xi}
  F.~Lizzi, S.~Vaidya and P.~Vitale,
  ``Twisted conformal symmetry in noncommutative two-dimensional quantum field
  theory,''
  Phys.\ Rev.\  D {\bf 73}, 125020 (2006)
  [arXiv:hep-th/0601056].


\bibitem{Tureanu:2006pb}
  A.~Tureanu,
  ``Twist and spin-statistics relation in noncommutative quantum field
  theory,''
  Phys.\ Lett.\  B {\bf 638}, 296 (2006)
  [arXiv:hep-th/0603219].
  
\bibitem{Zahn:2006wt}
  J.~Zahn,
  ``Remarks on twisted noncommutative quantum field theory,''
  Phys.\ Rev.\  D {\bf 73}, 105005 (2006)
  [arXiv:hep-th/0603231].

  
\bibitem{Bu:2006ha}
  J.~G.~Bu, H.~C.~Kim, Y.~Lee, C.~H.~Vac and J.~H.~Yee,
  ``Noncommutative field theory from twisted Fock space,''
  Phys.\ Rev.\  D {\bf 73}, 125001 (2006)
  [arXiv:hep-th/0603251].

\bibitem{Abe:2006ig}
  Y.~Abe,
  ``Noncommutative quantization for noncommutative field theory,''
  arXiv:hep-th/0606183.

\bibitem{Balachandran:2006pi}
  A.~P.~Balachandran, T.~R.~Govindarajan, G.~Mangano, A.~Pinzul, B.~A.~Qureshi and S.~Vaidya,
  ``Statistics and UV-IR mixing with twisted Poincare invariance,''
  Phys.\ Rev.\  D {\bf 75}, 045009 (2007)
  [arXiv:hep-th/0608179].

\bibitem{Fiore:2007vg}
  G.~Fiore and J.~Wess,
  ``On 'full' twisted Poincare symmetry and QFT on Moyal-Weyl spaces,''
  arXiv:hep-th/0701078.

\bibitem{Joung:2007qv}
  E.~Joung and J.~Mourad,
  ``QFT with twisted Poincare invariance and the Moyal product,''
  arXiv:hep-th/0703245.
  
  
\bibitem{Freidel:2005bb}
  L.~Freidel and E.~R.~Livine,
  ``Ponzano-Regge model revisited. III: Feynman diagrams and effective  field
  theory,''
  Class.\ Quant.\ Grav.\  {\bf 23}, 2021 (2006)
  [arXiv:hep-th/0502106].
  
\bibitem{Noui:2006kv}
  K.~Noui,
  ``Three dimensional loop quantum gravity: Towards a self-gravitating quantum
  field theory,''
  Class.\ Quant.\ Grav.\  {\bf 24}, 329 (2007)
  [arXiv:gr-qc/0612145].
  
\bibitem{Noui:2006ku}
  K.~Noui,
  ``Three dimensional loop quantum gravity: Particles and the quantum double,''
  J.\ Math.\ Phys.\  {\bf 47}, 102501 (2006)
  [arXiv:gr-qc/0612144].
  

\bibitem{Oeckl:1999zu}
  R.~Oeckl,
  ``Braided quantum field theory,''
  Commun.\ Math.\ Phys.\  {\bf 217}, 451 (2001)
  [arXiv:hep-th/9906225].
  
\bibitem{Oeckl:2000eg}
  R.~Oeckl,
  ``Untwisting noncommutative R**d and the equivalence of quantum field
  theories,''
  Nucl.\ Phys.\  B {\bf 581}, 559 (2000)
  [arXiv:hep-th/0003018].


\bibitem{Freidel:2006gc}
  L.~Freidel, J.~Kowalski-Glikman and S.~Nowak,
  ``From noncommutative kappa-Minkowski to Minkowski space-time,''
  arXiv:hep-th/0612170.

  
\bibitem{Majid:1996kd}
  S.~Majid,
  ``Foundations of quantum group theory,''
{\it  Cambridge, UK: Univ. Pr. (1995) 607 p}

\bibitem{Majid:1992sn}
  S.~Majid,
  ``Beyond supersymmetry and quantum symmetry: An Introduction to braided
  groups and braided matrices,''
  arXiv:hep-th/9212151.

\bibitem{Klimyk:1997eb}
  A.~Klimyk and K.~Schmudgen,
  ``Quantum groups and their representations,''
{\it  Berlin, Germany: Springer (1997) 552 p}

\bibitem{Ponzano}
G.~Ponzano and T.~Regge,
in ``Spectroscopic and Group Theoretical Methods in Physics''
ed. F.~Bloch, {\it North-Holland, Amsterdam, (1968)}. 

\bibitem{Alvarez-Gaume:2006bn}
  L.~Alvarez-Gaume, F.~Meyer and M.~A.~Vazquez-Mozo,
  ``Comments on noncommutative gravity,''
  Nucl.\ Phys.\  B {\bf 753}, 92 (2006)
  [arXiv:hep-th/0605113].

\bibitem{Agostini:2006nc}
  A.~Agostini, G.~Amelino-Camelia, M.~Arzano, A.~Marciano and R.~A.~Tacchi,
  ``Generalizing the Noether theorem for Hopf-algebra spacetime symmetries,''
  arXiv:hep-th/0607221.
  
\bibitem{Szabo:2001kg}
  R.~J.~Szabo,
  ``Quantum field theory on noncommutative spaces,''
  Phys.\ Rept.\  {\bf 378}, 207 (2003)
  [arXiv:hep-th/0109162].

\end{thebibliography}
\end{document}